# Fast Training Algorithms for Deep Convolutional Fuzzy Systems with Application to Stock Index Prediction

Li-Xin Wang

*Abstract*—**A deep convolutional fuzzy system (DCFS) on a high-dimensional input space is a multi-layer connection of many low-dimensional fuzzy systems, where the input variables to the low-dimensional fuzzy systems are selected through a moving window across the input spaces of the layers. To design the DCFS based on input-output data pairs, we propose a bottom-up layer-by-layer scheme. Specifically, by viewing each of the first-layer fuzzy systems as a weak estimator of the output based only on a very small portion of the input variables, we design these fuzzy systems using the WM Method. After the first-layer fuzzy systems are designed, we pass the data through the first layer to form a new data set and design the second-layer fuzzy systems based on this new data set in the same way as designing the first-layer fuzzy systems. Repeating this process layer-by-layer we design the whole DCFS. We also propose a DCFS with parameter sharing to save memory and computation. We apply the DCFS models to predict a synthetic chaotic plus random time-series and the real Hang Seng Index of the Hong Kong stock market.**

## I. INTRODUCTION

The great success of deep convolutional neural networks (DCNN) [1,2] in solving complex practical problems [3,4] reveals a basic fact that multi-level structures are very powerful models in representing complex relationships. The main problems of DCNN are the huge computational load to train the tones of parameters of the DCNN and the lack of interpretability for these huge number of model parameters [5]. The goal of this paper is to develop *deep convolutional fuzzy systems* (DCFS) and fast training algorithms for the DCFS to explore the power of multi-level rule-based representations and to overcome the computational and interpretability difficulties of DCNN.

Hierarchical fuzzy systems were proposed by Raju, Zhou and Kisner [6] in 1991, roughly the same time when LeCun [1] introduced the deep convolutional neural networks in 1990. In the late 1990s, the basic properties of hierarchical fuzzy systems such as universal approximation were proved in [7] and a back-propagation algorithm was developed in [8] to train the hierarchical fuzzy systems based on input-output data. Then, a wave of research on hierarchical fuzzy systems was conducted in the fuzzy community around the middle 2000s -- the same period when Hinton [2] proposed the celebrated new training algorithm for deep neural networks in 2006 which has led to the current AI boom. During this period, the structural and approximation properties of hierarchical fuzzy systems were studied in depth [9,10,11,12] and many new methods for designing hierarchical fuzzy systems were proposed [13,14,15, 16]. Since then, hierarchical fuzzy models have been applied to a wide variety of practical problems, such as environmental monitoring [17], educational assessment [18], video de-interlacing [19], price negotiation [20], mobile robots automation [21,22,23], self-nominating in peer-to-peer networks [24], linguistic hierarchy [25], hotel location selection [26], smart structures [27], weapon target assignment [28], image description [29], nutrition evaluation [30], spacecraft control [31], photovoltaic management [32], wastewater treatment [33], etc.. More recently, the research on hierarchical fuzzy systems has been advanced along many directions, such as fast implementation [34], adaptive control [35], multi-objective optimization [36], interpretability [37], classification [38,39], etc..

The researches summarized in the last paragraph demonstrate that the hierarchical fuzzy models are successful in solving many practical problems. However, the applications of hierarchical fuzzy models have been generally restricted to low dimensional problems with small data sets. Furthermore, the current training algorithms for hierarchical fuzzy systems are of the same types as those for deep neural networks, which are computationally intensive when applied to high-dimensional problems with big data. The heavy computational load is mainly due to the iterative nature of the training algorithms (multiple passes of the data) and it may take a long time to converge. Since the parameters of fuzzy systems have clear physical meanings (a clear connection to the input/output variables and the data) which the neural network parameters do not have, we can take advantage of these physical meanings to develop fast training algorithms for the parameters. The Wang-Mendel (WM) Method proposed in [40,41] is such a fast training algorithm that uses the training data only one-pass to determine the fuzzy system parameters.

The basic idea of this paper is to use the WM Method to design the low-dimensional fuzzy systems in a bottom-up, layer-by-layer fashion so that a deep convolutional fuzzy system is eventually constructed, where the inputs to the low-dimensional fuzzy systems are selected through a convolutional operator (a moving window). These low-dimensional fuzzy systems may be viewed as weak

Li-Xin Wang is with the University of Chinese Academy of Sciences, Beijing, P.R. China (e-mail: lxwang@ucas.edu.cn).

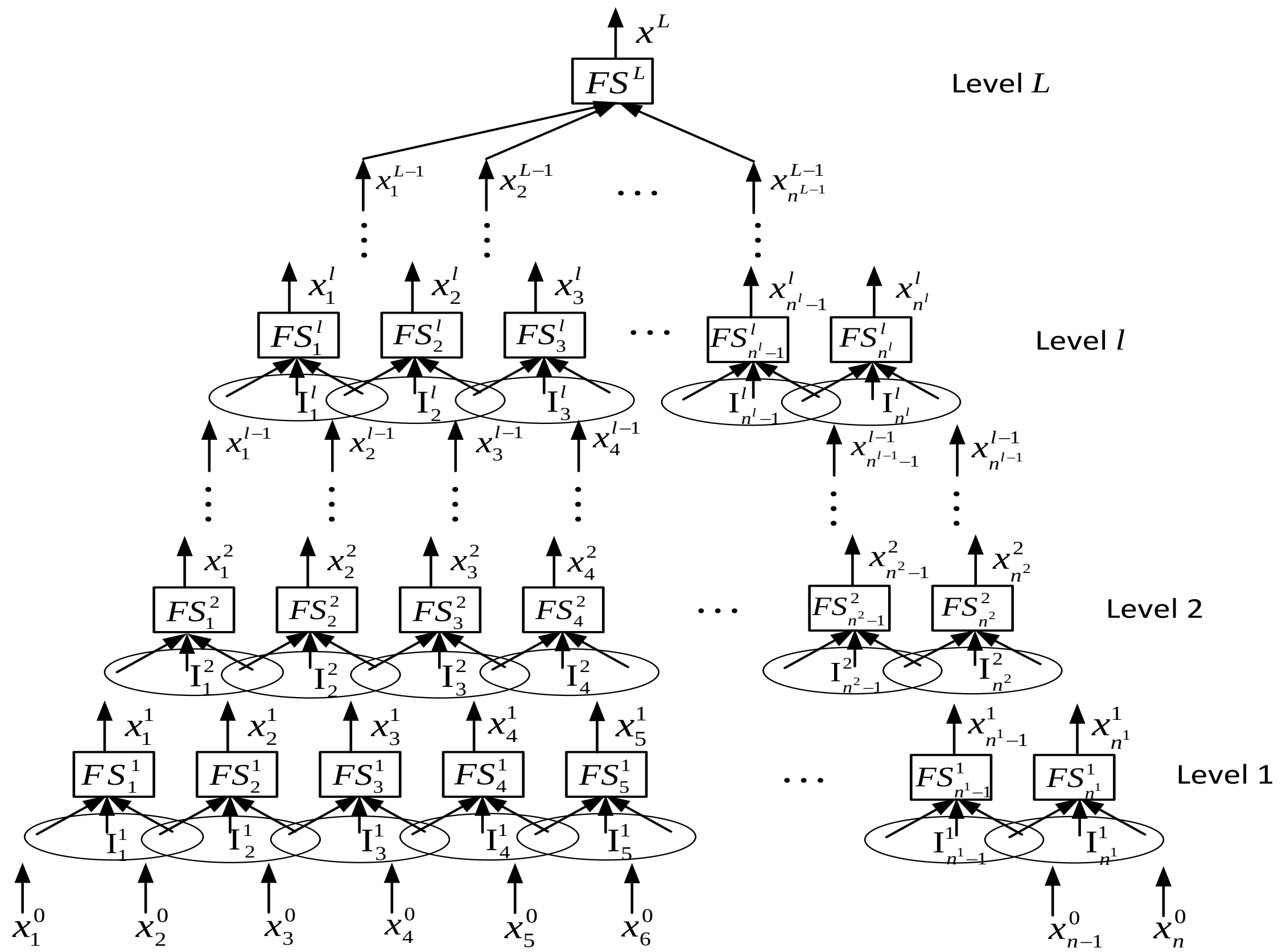

Fig. 1: The general structure of a deep convolutional fuzzy system (DCFS).

estimators [42] of the output variable. But, unlike the classical ensemble methods in machine learning [43] such as bagging [44], random forest [45] or boosting [46], the weak estimators (the low-dimensional fuzzy systems) in our DCFS models are constructed in a layer-by-layer fashion. Specifically, the first-level fuzzy systems may be viewed as the ordinary weak estimators, with each fuzzy system uses only a very small number of the input variables from the high-dimensional input space. After the first-level fuzzy systems are designed using the standard WM Method, they are fixed and their outputs form the input space to the second-level fuzzy systems. By passing the training data through the fixed first-level fuzzy systems, a new data set is generated and the second-level fuzzy systems are designed based on this new data set in the same way as the first-level fuzzy systems. This process continues, layer after layer, until the DCFS is constructed.

To test the DCFS models and the training algorithms, we apply them to predict a synthetic chaotic plus random time-series and the real Hang Seng Index of the Hong Kong stock market. Although it was generally believed that stock prices follow random walks [47,48] and therefore are not predictable, many researches showed that the stock prices do not follow random walks [49,50] and demonstrated that predicting the market is possible [51,52,53,54,55]. Since the stock prices are driven by the buying and selling operations of human traders who are influenced by human psychology such as greed and fear [56,57], it is reasonable to believe that there are some predictable elements in the stock prices. Of course, it is a very challenging task to catch up with these predictable elements in a timely fashion to make a profit [53,54].

This paper is organized as follows. In Section II, we show the structural details of the DCFS. In Section III, we develop four training algorithms for the DCFS. In Section IV, we apply the DCFS models with the training algorithms to predict a chaotic plus random time-series and the real Hang Seng Index of the Hong Kong stock market. Finally, some concluding remarks are drawn in Section V, and the MATLAB code of the main training algorithm is provided in the Supplemental Material.

## II. Structure of Deep Convolutional Fuzzy Systems

We begin with the definition of the general deep convolutional fuzzy system.

**Definition 1**: The general structure of a *deep convolutional fuzzy system* (DCFS) is illustrated in Fig. 1, where the input vector $(x_1^0, x_2^0, \ldots, x_n^0)$ to the DCFS is generally of very high dimension, and the output $x^L$ is a scalar (a multi-output DCFS may be designed as multiple single-output DCFSs). Level $l$ ($l$=1,2, ..., $L$-1) consists of $n^l$ fuzzy systems $FS_i^l$ ($i$=1,2, ...,$n^l$) whose outputs are denoted as $x_i^l$ which are inputs to Level $l$+1. The top level, Level L, has only one fuzzy system $FS^L$ that combines the $n^{L-1}$outputs from Level L-1 to produce the final output $x^L$. The input sets $I_1^l$, $I_2^l$, …, $I_{n^l}^l$ to the fuzzy systems $FS_1^l$, $FS_2^l$, …, $FS_{n^l}^l$ ($l$=1,2, ..., $L$-1) are selected from the previous level's outputs $x_1^{l-1}$, $x_2^{l-1}$, …, $x_{n^{l-1}}^{l-1}$ through a moving window of length $m$, where the window size $m$ is usually a small number such as 3, 4 or 5. ■

The moving window may take a variety of moving schemes. For example, it may move one variable at a time starting from $x_1^{l-1}$ until $x_{n^{l-1}}^{l-1}$ is covered, and this gives

$$\begin{aligned} I_1^l &= (x_1^{l-1}, \dots, x_m^{l-1}), \\ I_2^l &= (x_2^{l-1}, \dots, x_{m+1}^{l-1}), \\ &\vdots \\ I_i^l &= (x_i^{l-1}, \dots, x_{m+i-1}^{l-1}), \\ &\vdots \\ I_{n^{l-1}-m+1}^l &= (x_{n^{l-1}-m+1}^{l-1}, \dots, x_{n^{l-1}}^{l-1}), \end{aligned} \tag{1}$$

where $l$=1,2, ..., $L$-1 (for Level $l$=1 we have $n^0 = n$). For this one-variable-at-a-time moving scheme, we have

$$n^l = n^{l-1} - m + 1 \tag{2}$$

for $l$=1,2, ..., $L$-1 with $n^0 = n$, from which we get

$$n^l = n - l(m-1). \tag{3}$$

If we do not want to use too many fuzzy systems $FS_i^l$ in the construction of the DCFS to improve the efficiency of each fuzzy system $FS_i^l$, we may move the window more than one variable at a time to cover the input variables in the levels. In the extreme case, we may move the window $m$ variables each time for all the fuzzy systems $FS_i^l$ so that a $L$ level DCFS can cover $m^L$ input variables. For a $L$=5 level DCFS with $m$=5 inputs to each fuzzy system $FS_i^l$, for example, $m^L = 3125$ input variables can be covered. Also, the window size $m$ may be different for different fuzzy systems $FS_i^l$ to introduce more flexibility to the DCFS model.

The fuzzy systems $FS_i^l$ ($i$=1,2, ...,$n^l$, $l$=1,2, ..., $L$-1) are standard fuzzy systems [58] constructed as follows. For each input variable $x_i^{l-1}, \dots, x_{m+i-1}^{l-1} \in I_i^l$ to the fuzzy system $FS_i^l$ (note that $x_i^{l-1}$ is the first input variable to $FS_i^l$ and may not be the $i'$th input variable to Level $l$), define q fuzzy sets $A^1$, $A^2$,..., $A^q$ as shown in Fig. 2, where the centers of the q fuzzy sets are equally spaced and the endpoints $minx_j$ and $maxx_j$ are determined from the training data (the details will be given in the next section when we develop the training algorithms for the DCFS). The fuzzy system $FS_i^l$: $(x_i^{l-1}, \dots, x_{m+i-1}^{l-1}) \to x_i^l$ is

$$\begin{aligned} x_i^l &= FS_i^l(x_i^{l-1}, \dots, x_{m+i-1}^{l-1}) \\ &= \frac{\sum_{j_1=1}^q \cdots \sum_{j_m=1}^q c^{j_1 \cdots j_m} A^{j_1}(x_i^{l-1}) \cdots A^{j_m}(x_{m+i-1}^{l-1})}{\sum_{j_1=1}^q \cdots \sum_{j_m=1}^q A^{j_1}(x_i^{l-1}) \cdots A^{j_m}(x_{m+i-1}^{l-1})}, \end{aligned} \tag{4}$$

which is constructed from the following $q^m$ fuzzy IF-THEN rules:

$$\text{IF } x_i^{l-1} \text{ is } A^{j_1} \text{ and} \dots \text{and } x_{m+i-1}^{l-1} \text{ is } A^{j_m}, \text{THEN y is } B^{j_1 \cdots j_m}, \tag{5}$$

where the membership functions $A^j$'s are given in Fig. 2, and the parameters $c^{j_1 \cdots j_m}$ are the centers of the fuzzy sets $B^{j_1 \cdots j_m}$ and will be designed using the training algorithms in the next section.

For the membership functions in Fig. 2 we see that $\sum_{j_1=1}^q A^{j_1}(x_i^{l-1}) = 1$, …, $\sum_{j_m=1}^q A^{j_m}(x_{m-i+1}^{l-1}) = 1$, so the denominator of the fuzzy system $FS_i^l$ of (4) is equal to 1:

$$\begin{aligned} &\sum_{j_1=1}^q \cdots \sum_{j_m=1}^q A^{j_1}(x_i^{l-1}) \cdots A^{j_m}(x_{m+i-1}^{l-1}) \\ &= \left(\sum_{j_1=1}^q A^{j_1}(x_i^{l-1})\right) \cdots \left(\sum_{j_1=1}^q A^{j_1}(x_i^{l-1})\right) \\ &= 1, \end{aligned} \tag{6}$$

and the fuzzy system $FS_i^l$ of (4) is simplified to

$$\begin{aligned} x_i^l &= FS_i^l(x_i^{l-1}, \dots, x_{m+i-1}^{l-1}) \\ &= \sum_{j_1=1}^q \cdots \sum_{j_m=1}^q c^{j_1 \cdots j_m} A^{j_1}(x_i^{l-1}) \cdots A^{j_m}(x_{m+i-1}^{l-1}), \end{aligned} \tag{7}$$

where $i$=1,2, ..., $n^l$ and $l$=1,2, ..., $L$-1. The top level fuzzy system $FS^L$ is in the same form of (7) with $n^{L-1}$ input variables $x_1^{L-1}$, …, $x_{n^{L-1}}^{L-1}$.

We claimed in the Introduction that the DCFS has better interpretability than the DCNN, and now we show the details of how to interpret the DCFS in terms of the fuzzy IF-THEN rules (5) and the parameters $c^{j_1 \cdots j_m}$. First, notice that the fuzzy system $FS_i^l(x_i^{l-1}, \dots, x_{m+i-1}^{l-1})$ of (7) is constructed from the $q^m$ fuzzy IF-THEN rules (5) with each rule covering a cell $(j_1, \cdots, j_m)$ in the $m$-dimensional input space to $FS_i^l(x_i^{l-1}, \dots, x_{m+i-1}^{l-1})$. Fig. 3 illustrates the case of $m$=2 and $q = 5$, where the 2-dimensional input space is partitioned into $q^m = 5^2 = 25$ cells with the cell $(i, j)$ $(i, j = 1, \dots, 5)$ covered by the fuzzy IF-THEN rule:

$$\text{IF } x_1 \text{ is } A^i \text{ and } x_2 \text{ is } A^j, \text{THEN y is } B^{ij}, \tag{8}$$

and the parameter $c^{ij}$ being the center of the fuzzy set $B^{ij}$. For a given input point, say the point $(x_1^*, x_2^*)$ in Fig. 3 (left), the action of the fuzzy system $FS(x_1^*, x_2^*)$ can be represented by the fuzzy IF-THEN rule: IF $x_1$ is $A^2$ and $x_2$ is $A^3$, THEN y is $B^{23}$, which covers the cell over $(x_1^*, x_2^*)$. That is, the fuzzy system takes local actions with one rule responsible mainly for one cell. Therefore, for a given point $(x_i^{l-1}, \dots, x_{m+i-1}^{l-1})$, the action of the fuzzy system $FS_i^l(x_i^{l-1}, \dots, x_{m+i-1}^{l-1})$ of (7) can be represented by a single parameter $c^{j_1 \cdots j_m}$ that represents the fuzzy IF-THEN rule in the form of (5).

Because each of the fuzzy system $FS_i^l$ in the general DCFS of Fig. 1 can be represented by a single parameter $c^{j_1 \cdots j_m}$ (as we demonstrated in the last paragraph), the whole action of the DCFS on any given input point can be interpreted by a graph of connected $c^{j_1 \cdots j_m}$'s. Fig. 3 (the right part) shows an example of 6-level DCFS with $n$=7 inputs to the DCFS and $m$=2 inputs to each fuzzy system $FS_i^l$ in the DCFS. For a given input $(x_1^*, \dots, x_7^*)$ in Fig. 3, if the DCFS gives a bad output $y$ (such as causing an accident in an auto-car application), then from the DCFS graph in the right part of Fig. 3 we can easily check what rules cause the bad output and take appropriate changes to these rules so that the mistake will not happen again. We will discuss this easy-error-correction property of DCFS in more details after we develop the training algorithms in the next section.

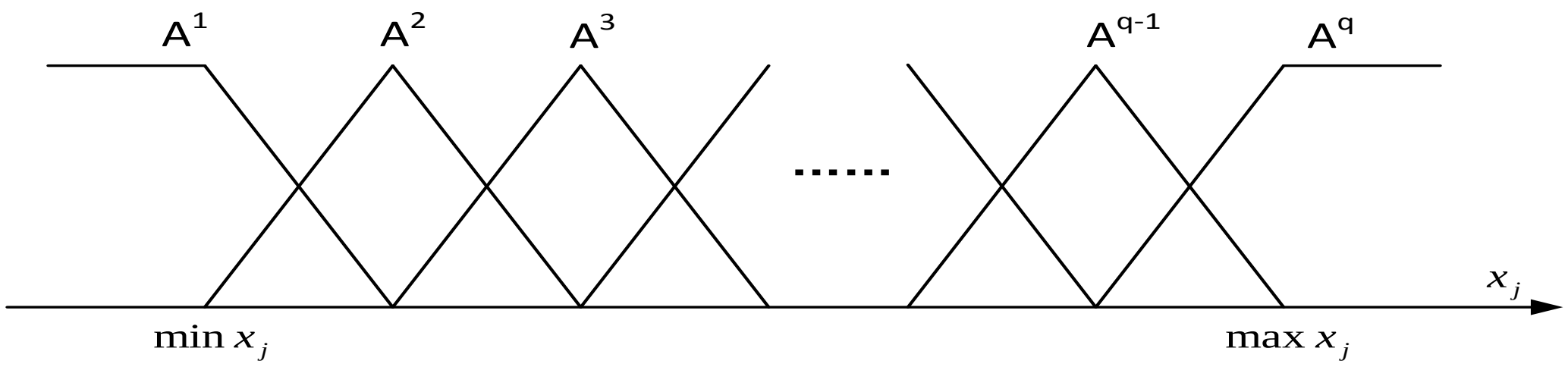

Fig. 2: The membership functions in the fuzzy systems $FS_i^l$ of the DCFS.

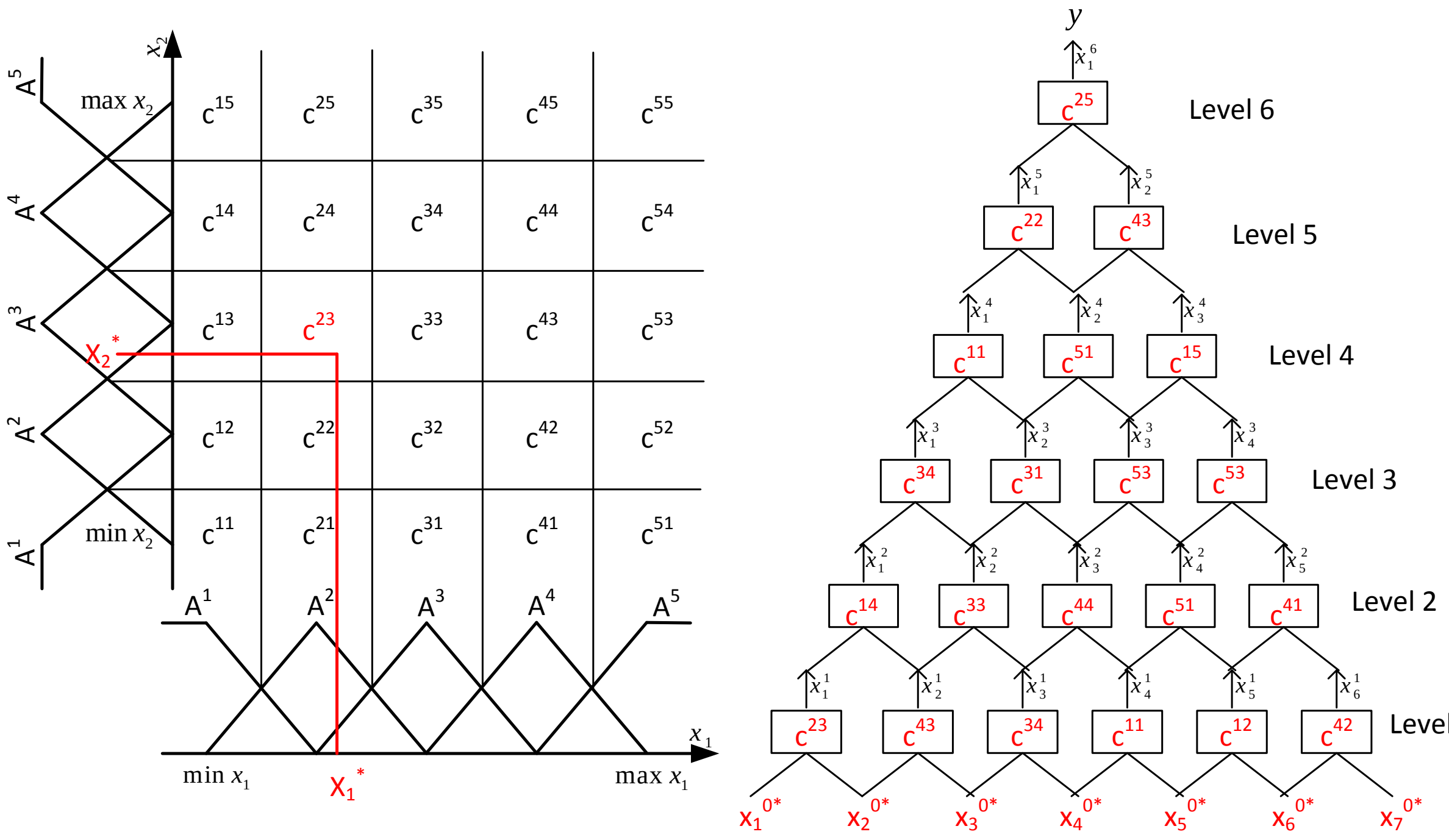

For given input point ($x_1^*, x_2^*$), the action of fuzzy system FS($x_1^*, x_2^*$) can be represented by a single parameter $c^{23}$

For a given input point ($x_1^{0*}, \cdots, x_7^{0*}$), a DCFS can be interpreted by a connection of the parameters $c^{ij}$

Fig. 3: Interpretation of DCFS. Left: The action of a fuzzy system on any given input point can be represented by the local fuzzy IF-THRN rule with parameter $c^{ij}$. Right: For a given input point, a DCFS can be interpreted by a networked connection of fuzzy IF-THEN rules with parameters $c^{ij}$.

Finally, we introduce a special DCFS that can greatly reduce the memory requirement and the computational cost of the training algorithms for the DCFS to be developed in the next section. We see from (7) that each fuzzy system $FS_i^l$ has $q^m$ free parameters $c^{j_1 \cdots j_m}$ to be designed and stored in the computer memory. The total computational and storage requirement is proportional to $q^m \sum_{l=1}^{L} n^l$, which may be a large number for high-dimensional problems. To reduce the computational and storage requirement, we introduce a *parameter sharing* scheme where the fuzzy systems in the same level are identical (sharing the same $c^{j_1 \cdots j_m}$ parameters). Specifically:

**Definition 2**: A *DCFS with parameter sharing* is a DCFS where the fuzzy systems $FS_i^l$ in the same level $l$ are identical, i.e., $FS_1^l = \cdots = FS_{n^l}^l$ for $l$=1,2, ..., $L$, and we denote this identical fuzzy system in Level $l$ as $FS^l$. A DCFS with parameter sharing has $q^m L$ free parameters $c^{j_1 \cdots j_m}$ to be designed and stored. ■

We now move to the next section to develop a number of fast training algorithms to determine the parameters $c^{j_1 \cdots j_m}$ of the fuzzy system (7) based on input-output data for the general DCFS and the DCFS with parameter sharing.

## III. Fast Training Algorithms for DCFS

**Task 1 (off-line training):** Given $N$ input-output data pairs:

$$[x_1^0(k), x_2^0(k), \dots, x_n^0(k); y^0(k)], \quad k = 1,2, \dots, N, \tag{9}$$

where $x_1^0(k), x_2^0(k), \dots, x_n^0(k)$ are the inputs and $y^0(k)$ is the output, our task is to design a DCFS in Fig. 1 to match these input-output data pairs. ■

First, we develop a training algorithm for the general DCFS in Definition 1 to match the input-output data pairs of (9) (Training Algorithm 1 below). Then, we show how to design the DCFS with parameter sharing in Definition 2 in Training Algorithm 2. Finally, in Training Algorithms 3 and 4 we show how to do on-line training for the general DCFS and the DCFS with parameter sharing, respectively.

**Training Algorithm 1 (for general DCFS):** Given the input-output data pairs of (9), we design the general DCFS in Definition 1 with the fuzzy systems $FS_i^l$ in the form of (7) through the following steps:

**Step 1:** Determine the structure of the DCFS. Specifically:

**1.1**: Choose the moving window size $m$ and the moving scheme (such as moving one-variable-at-a-time or other schemes).

**1.2**: Determine the number of levels $L$.

**Step 2:** Design the Level 1 fuzzy systems $FS_i^1$ in the form of (7) (with $m$ input $x_i^1, \dots, x_{m+i-1}^1$), $i$=1,2, ..., $n^1$, using the WM Method [40,41], where the input-output data pairs used to design the $FS_i^1$ are:

$$[x_i^0(k), \dots, x_{m+i-1}^0(k); y^0(k)], \ \ k = 1,2, \dots, N, \tag{10}$$

Specifically:

**2.1**: For each cell $(j_1, \cdots, j_m)$ with $j_1, \cdots, j_m = 1,2, \dots, \mathrm{q}$, set the initial values of the weight parameter $w^{j_1,\cdots,j_m}$ and the weight-output parameter $u^{j_1,\cdots,j_m}$ equal to zero.

**2.2**: For each input $x_i^0, \dots, x_{m+i-1}^0$ to $FS_i^1$, consider the q fuzzy sets $A^1$, $A^2$,..., $A^q$ in Fig. 2 and choose the endpoints as

$$\begin{aligned} minx_i^0 &= min(x_i^0(k) \,|\, k = 1,2, \dots, N), \\ maxx_i^0 &= max(x_i^0(k) \,|\, k = 1,2, \dots, N), \\ &\vdots \\ minx_{m+i-1}^0 &= min(x_{m+i-1}^0(k) \,|\, k = 1,2, \dots, N), \\ maxx_{m+i-1}^0 &= max(x_{m+i-1}^0(k) \,|\, k = 1,2, \dots, N). \end{aligned} \tag{11}$$

**2.3**: For each input-output data pair of (10) starting from $k = 1$, determine the fuzzy sets $A^{j_1^*}, \dots, A^{j_m^*}$ which achieve the maximum membership values among the q fuzzy sets $A^1$, $A^2$, ..., $A^q$ at $x_i^0(k), \dots, x_{m+i-1}^0(k)$, respectively, i.e., determine

$$\begin{aligned} j_1^* &= \arg\max\nolimits_{j\in\{1,2,\dots,\mathrm{q}\}} \left(A^j\left(x_i^0(k)\right)\right), \\ &\vdots \\ j_m^* &= \arg\max\nolimits_{j\in\{1,2,\dots,\mathrm{q}\}} \left(A^j\left(x_{m+i-1}^0(k)\right)\right). \end{aligned} \tag{12}$$

**2.4**: Update the weight and weight-output parameters for cell $(j_1^*, \cdots, j_m^*)$, $w^{j_1^*,\cdots,j_m^*}$ and $u^{j_1^*,\cdots,j_m^*}$, through

$$w^{j_1^*,\cdots,j_m^*} = w^{j_1^*,\cdots,j_m^*} + A^{j_1^*}\left(x_i^0(k)\right) \cdots A^{j_m^*}\left(x_{m+i-1}^0(k)\right), \tag{13}$$

$$u^{j_1^*,\cdots,j_m^*} = u^{j_1^*,\cdots,j_m^*} + \left(A^{j_1^*}(x_i^0(k)) \cdots A^{j_m^*}(x_{m+i-1}^0(k))\right) y^0(k). \tag{14}$$

**2.5**: Repeat 2.3 and 2.4 for $k = 1,2, \dots, N$. For the cells $(j_1, \cdots, j_m)$ with $w^{j_1\cdots j_m} \neq 0$, determine the parameters $c^{j_1\cdots j_m}$ in the fuzzy system $FS_i^1$ of (7) as

$$c^{j_1\cdots j_m} = {u^{j_1\cdots j_m}}/{w^{j_1\cdots j_m}}\ . \tag{15}$$

We call the cells $(j_1, \cdots, j_m)$ with $w^{j_1\cdots j_m} \neq 0$ *covered by data*, and define

$$\mathrm{C}(0) = \{(j_1, \cdots, j_m) \mid (j_1, \cdots, j_m)\ is\ covered\ by\ data\} \tag{16}$$

**2.6:** For each cell $(j_1, \cdots, j_m)$ not in $\mathrm{C}(0)$, search its neighbors to see whether they are in $\mathrm{C}(0)$, where two cells $(j_1, \cdots, j_m)$ and $(j_1', \cdots, j_m')$ are *neighbors to each other* if $j_i = j_i'$ for all $i = 1, \dots, m$ except at one location $r$ such that $j_r = j_r' + 1$ or $j_r = j_r' - 1$. For the cells $(j_1, \cdots, j_m)$ not in $\mathrm{C}(0)$ that have at least one neighbor in $\mathrm{C}(0)$, determine the $c^{j_1\cdots j_m}$ as the average of the $c^{j_1\cdots j_m}$'s of its neighbors in $\mathrm{C}(0)$. Define

$$\mathrm{C}(1) = \mathrm{C}(0) \oplus \{(j_1, \cdots, j_m) \mid c^{j_1\cdots j_m}\ determined\ in\ this\ step\} \tag{17}$$

**2.7:** Repeat 2.6 with $\mathrm{C}(0)$ replaced by $\mathrm{C}(1)$ and $\mathrm{C}(1)$ replaced by $\mathrm{C}(2)$, and continue this process to get $\mathrm{C}(3)$, $\mathrm{C}(4)$, …, until $\mathrm{C}(p)$ which contains all the cells $(j_1, \cdots, j_m)$ with $j_1, \cdots, j_m = 1,2, \dots, \mathrm{q}$. Fig. 4 illustrates this process. (More details of the WM Method are given in [41].)

**Step 3:** Suppose the fuzzy systems in Level 1 to Level $l$-1 have already been designed, we now design the fuzzy systems $FS_i^l$ , $i$=1,2, ..., $n^l$, in Level $l$, starting from $l$=2 (in this case the Level $l$-1=1 fuzzy systems $FS_i^1$, $i$=1,2, ..., $n^1$, have already been designed in Step 2). Specifically:

**3.1**: For each input-output data pair of (9) with $k = 1,2, \dots, N$, put $x_1^0(k), x_2^0(k), \dots, x_n^0(k)$ as inputs to Level 1 and compute upwards along the DCFS to get the outputs of Level $l$-1, denoted as $x_1^{l-1}(k), x_2^{l-1}(k), \dots, x_{n^{l-1}}^{l-1}(k)$; view

$$\left[x_1^{l-1}(k), x_2^{l-1}(k), \dots, x_{n^{l-1}}^{l-1}(k); y^0(k)\right], \ \ k = 1,2, \dots, N, \tag{18}$$

as the new input-output data pairs for designing the Level $l$ fuzzy systems $FS_i^l$.

**3.2**: Use the same procedure in Step 2 to design the Level $l$ fuzzy systems $FS_i^l$ in the form of (7) , $i$=1,2, ..., $n^l$, with the original input-output data pairs (9) replaced by the new input-output data pairs (18), and all the Level 1 variables in Step 2 replaced by the corresponding Level $l$ variables (for example, replace $n^1$ by $n^l$ , $x_i^0, \dots, x_{m+i-1}^0$ by $x_i^{l-1}, \dots, x_{m+i-1}^{l-1}$, etc.).

**3.3**: Set $l$=$l$+1 and repeat 3.1 and 3.2 until the Level $L$ fuzzy system $FS^L$ is designed.

■

Figs. 5 to 7 illustrate Training Algorithm 1 for the design of a three-level DCFS with 15 inputs and two-variable-at-a-time moving window scheme, where Figs. 5 to 7 are for the design of Levels 1 to 3, respectively. The MATLAB code of Training Algorithm 1 for a 11-input 5-level DCFS is given in the Supplemental Material, which can be easily modified for other applications. A few remarks about Training Algorithm 1 are now in order.

**Remark 1 (layer-by-layer construction)**: As discussed in the Introduction, the DCFS with Training Algorithm 1 may be viewed as a layered combination of weak estimators. Specifically, among the high-dimensional input variables $x_1^0, x_2^0, \dots, x_n^0$ to the DCFS, each fuzzy system $FS_i^1$ ($i$=1,2, ..., $n^1$) in Level 1 selects only a very small number of $m$ variables as its inputs, so each fuzzy system $FS_i^1$ may be viewed as a weak estimator for the output $y^0$. After these $n^1$ Level 1 weak estimators $FS_i^1$ are designed, they are fixed and their outputs $x_1^1, x_2^1, \dots, x_{n^1}^1$ form the $n^1$–dimensional input space to the Level 2 fuzzy systems $FS_i^2$ ($i$=1,2, …, $n^2$). Here again as in Level 1, each Level 2 fuzzy system $FS_i^2$ selects only a small number of $m$ variables from $x_1^1, x_2^1, \dots, x_{n^1}^1$ as its inputs, and the $n^2$ fuzzy systems $FS_i^2$ ($i$=1,2, …, $n^2$) are viewed as Level 2 weak estimators for the output $y^0$. This process continues up to the top Level L whose output $x^L$ is the final estimate of the $y^0$.

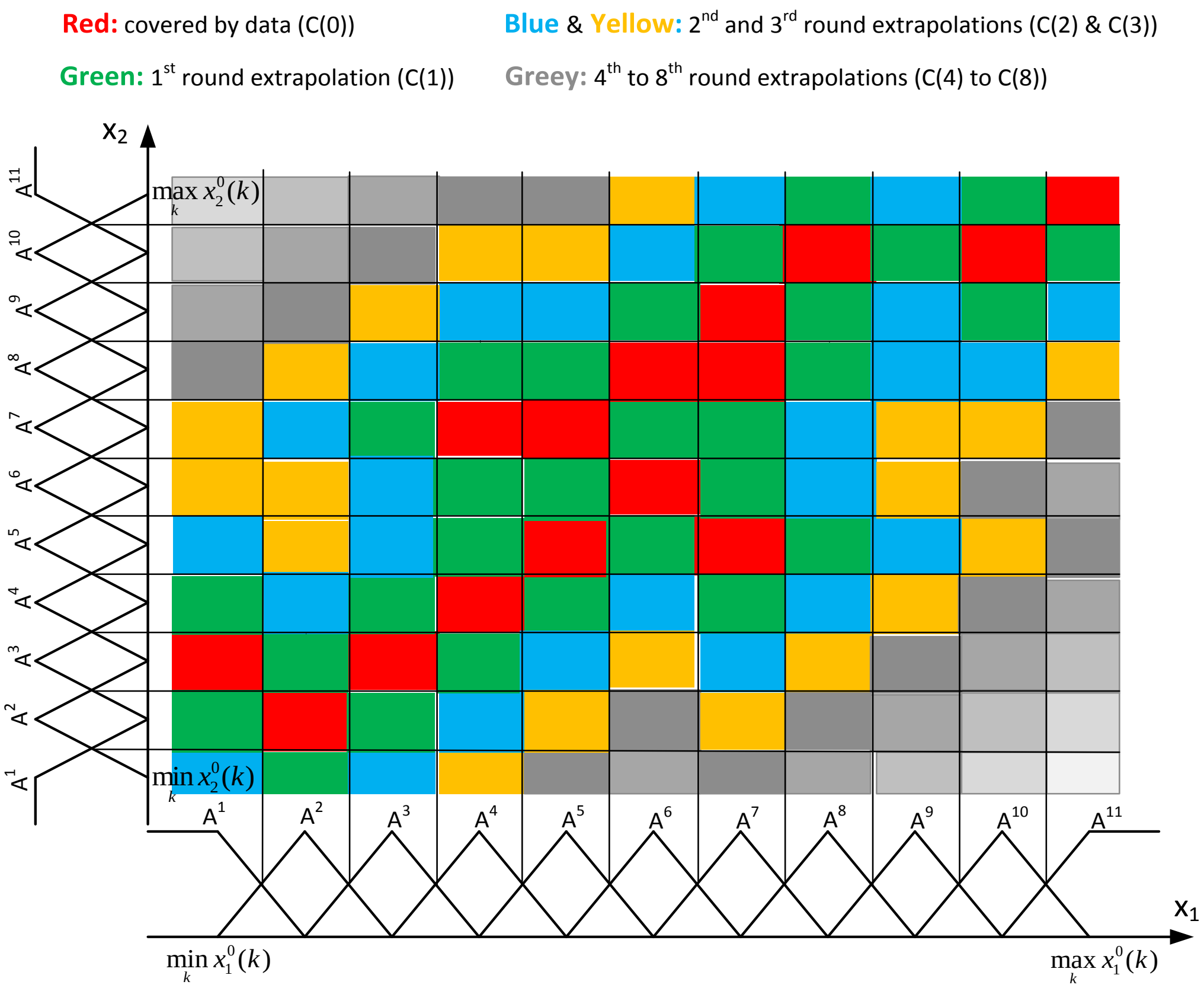

Fig. 4: Illustration of the WM Method to expend the cells from those covered by data (red) to the whole input space, where $\mathrm{C}(0)$ = red, $\mathrm{C}(1) = \mathrm{C}(0)$ + green, $\mathrm{C}(2) = \mathrm{C}(1)$ + blue, $\mathrm{C}(3) = \mathrm{C}(2)$ + yellow, …, and $\mathrm{C}(8) =$ all the cells.

**Remark 2 (physical meaning of the parameter design)**: From (13)-(15) we see that the parameters $c^{j_1 \cdots j_m}$ in the fuzzy system (7) are designed as the weighted average of the outputs $y^0(k)$ whose corresponding inputs $x_i^{l-1}(k), \dots, x_{m+i-1}^{l-1}(k)$ fall into the cell $(j_1 \cdots j_m)$, with the weight equal to the membership value $A^{j_1}(x_i^{l-1}(k)) \cdots A^{j_m}(x_{m+i-1}^{l-1}(k))$. If no data falls into a cell $(j_1 \cdots j_m)$, then the $c^{j_1 \cdots j_m}$ of the cell is determined through the extrapolation scheme of Steps 2.6 and 2.7 (illustrated in Fig. 4). We can use this simple scheme to design the parameters $c^{j_1 \cdots j_m}$ because of the clear physical meaning of the $c^{j_1 \cdots j_m}$. Specifically, the $c^{j_1 \cdots j_m}$ is the center of the THEN part membership function of the fuzzy IF-THEN rule (5) for the cell $(j_1 \cdots j_m)$, so the $c^{j_1 \cdots j_m}$ may be viewed as the estimate of the desired output $y^0$ based on the fuzzy IF-THEN rule (5) at cell $(j_1 \cdots j_m)$. Therefore, a good way to design the $c^{j_1 \cdots j_m}$ is to put it as the weighted average of the $y^0(k)$'s whose corresponding inputs $x_i^{l-1}(k), \dots, x_{m+i-1}^{l-1}(k)$ fall into the cell $(j_1 \cdots j_m)$. The optimality of designing the parameters in this way is studied in [59].

**Remark 3 (fast training with low computational cost)**: From 2.3 to 2.5 in Training Algorithm 1 we see that to design each of the fuzzy systems $FS_i^l$ in the DCFS, the $N$ input-output data pairs (10) are used only once (just one-pass through the data). In the popular gradient-decent-based back-propagation algorithm, multiple passes through the data are needed to ensure the convergence of the parameters, so the computational cost is high and the speed of the algorithm is slow. Since the data are passed through just once in Training Algorithm 1, it is a very fast algorithm. Specifically, the computational load of Training Algorithm 1 (for general DCFS) is approximately $O(N + q^m)\sum_{l=1}^{L} n^l$, where $O(N)$ accounts for the one-pass of data in the computation of (12)-(14), $O(q^m)$ accounts for the computation of the parameters $c^{j_1 \cdots j_m}$ in Steps 2.5-2.7 (illustrated in Fig. 4), and $\sum_{l=1}^{L} n^l$ is the number of fuzzy systems $FS_i^l$ ($i$=1,2, …,$n^l$, $l$=1,2, …, $L$) in the DCFS.

Since each fuzzy system $FS_i^l$ (7) in the general DCFS has $q^m$ free parameters $c^{j_1 \cdots j_m}$, the total computational and storage requirement is proportional to $q^m \sum_{l=1}^{L} n^l$, which may be a large number for high-dimensional problems. Therefore, we introduced the DCFS with parameter sharing in Definition 2. Now we show how to design the DCFS with parameter sharing based on the input-output data pairs (9).

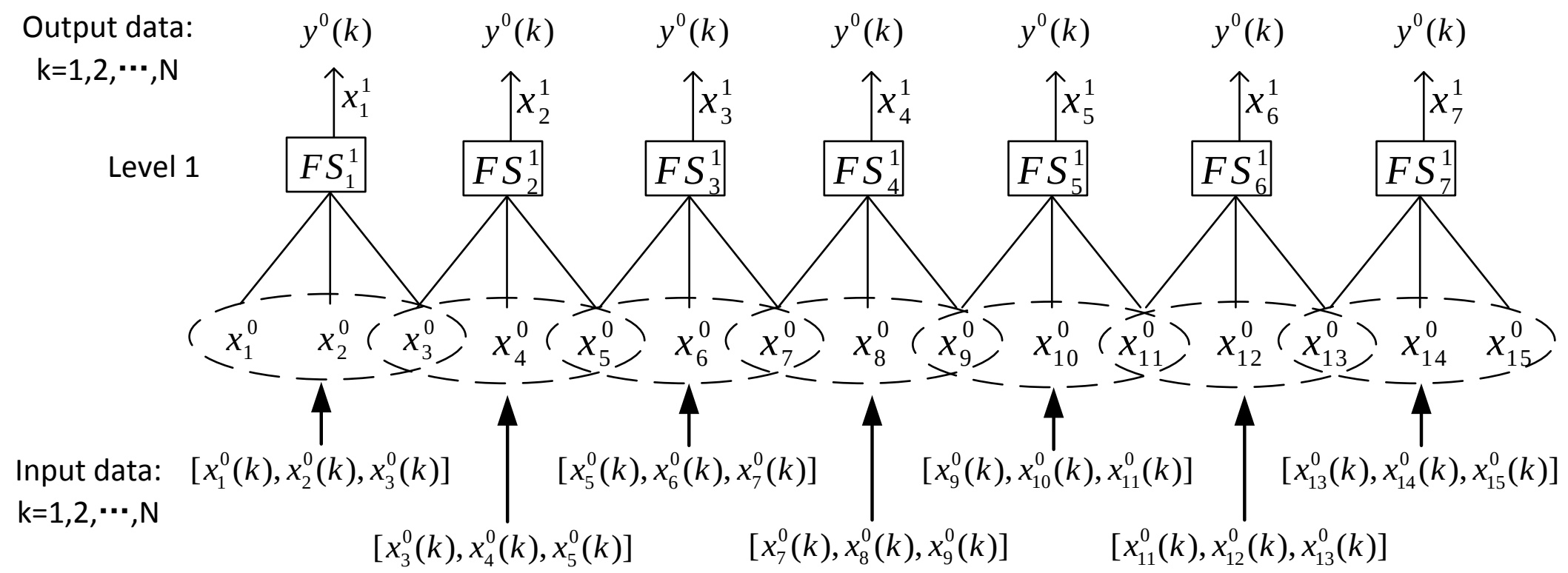

Fig. 5: Illustration of Training Algorithm 1 for general DCFS in the design of the Level 1 fuzzy systems $FS_i^1$.

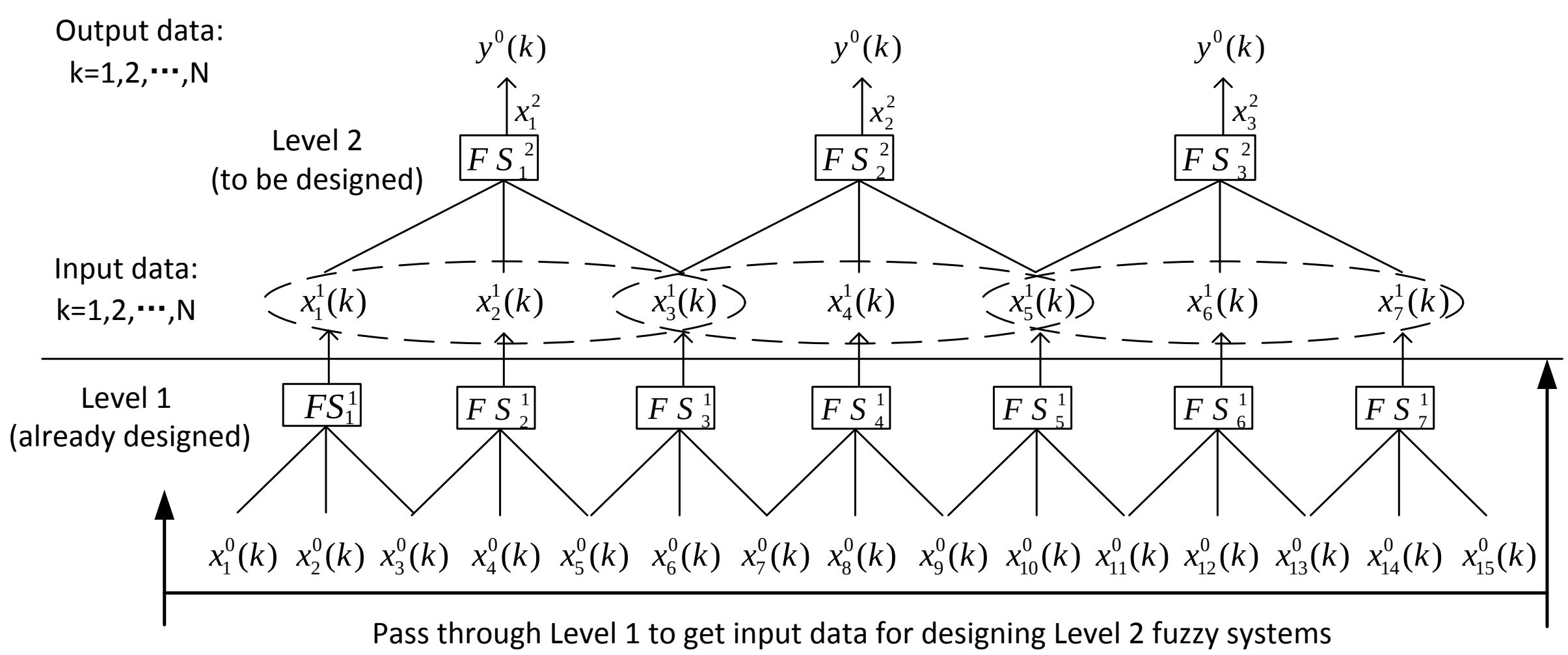

Fig. 6: Illustration of Training Algorithm 1 for general DCFS in the design of the Level 2 fuzzy systems $FS_i^2$.

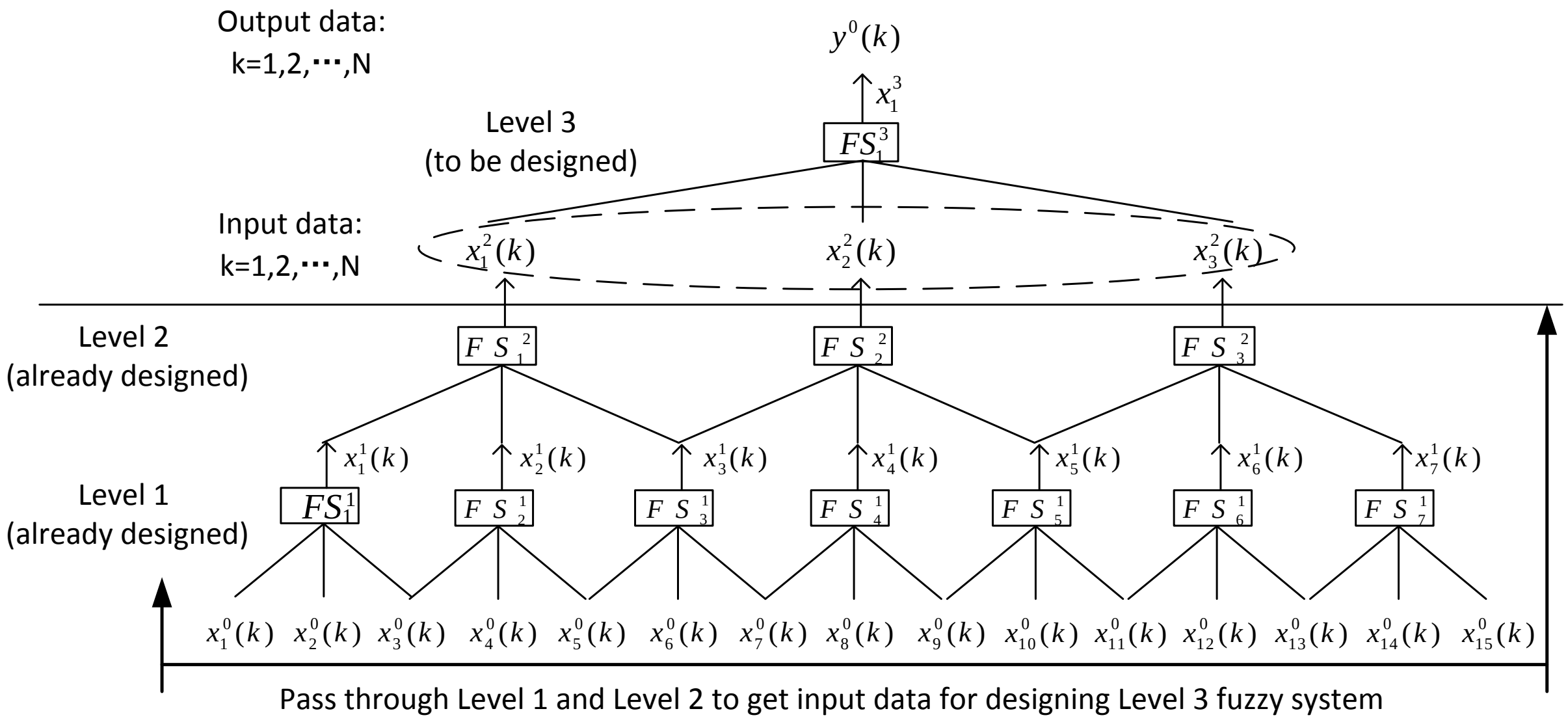

Fig. 7: Illustration of Training Algorithm 1 for general DCFS in the design of the Level 3 fuzzy system $FS^3$.

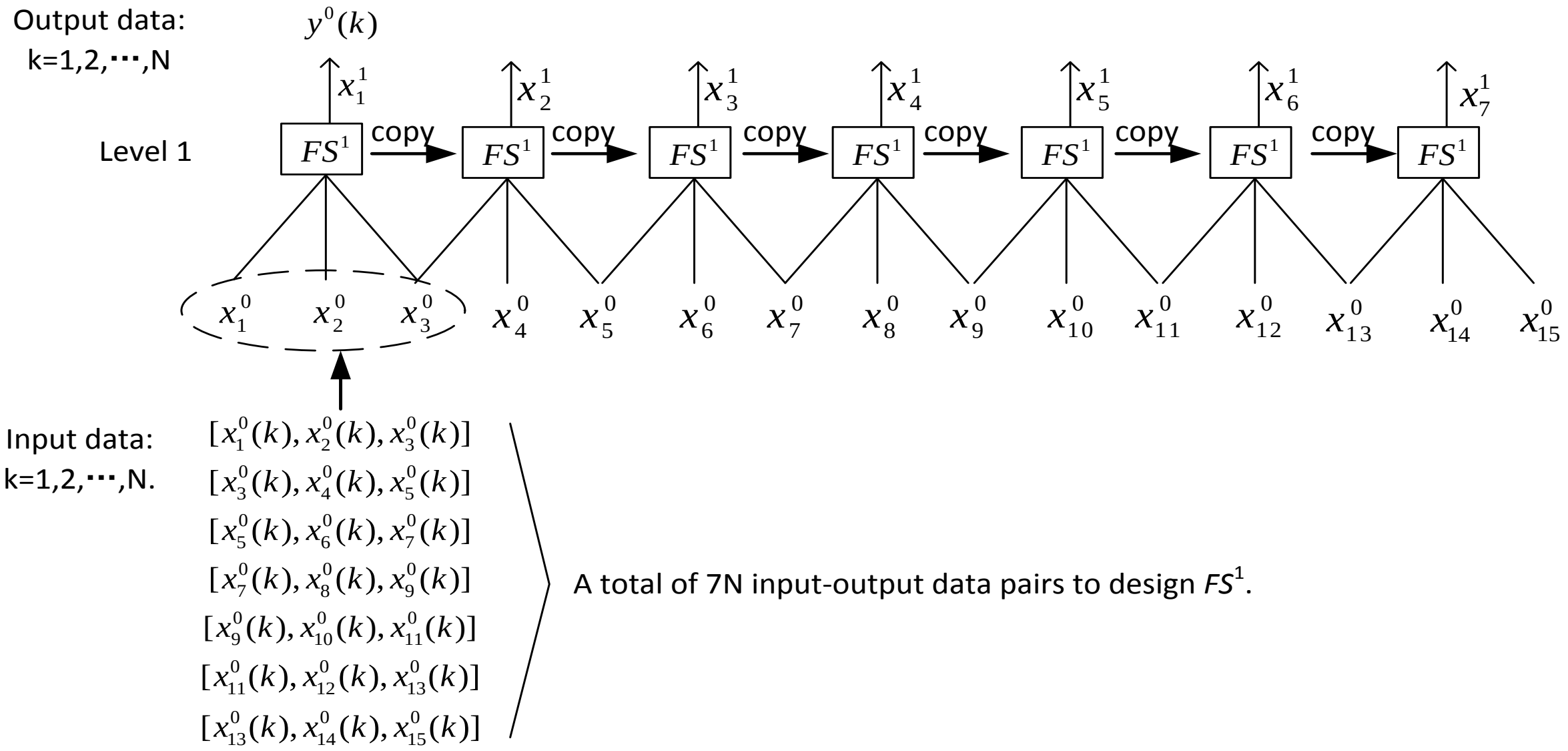

Fig. 8: Illustration of Training Algorithm 2 for DCFS with parameter sharing in the design of the Level 1 fuzzy systems $FS_1^1 = \cdots = FS_7^1 = FS^1$.

**Training Algorithm 2 (for DCFS with parameter sharing):** Given the input-output data pairs of (9), the DCFS with parameter sharing in Definition 2 are designed as follows:

**Step 1:** The same as Step 1 in Training Algorithm 1.

**Step 2:** Design the first Level 1 fuzzy system $FS_1^1 = FS^1$ in the form of (7) with the following $n^1 N$ input-output data pairs:

$$\begin{gathered}[x_1^0(k), \ldots, x_m^0(k); y^0(k)], \quad k = 1,2, \ldots, N, \\ \vdots \\ \left[x_{n^1}^0(k), \ldots, x_{m+n^1-1}^0(k); y^0(k)\right], \quad k = 1,2, \ldots, N,\end{gathered} \tag{19}$$

using the same procedure **2.1** to **2.7** in Training Algorithm 1. Then, copy this $FS_1^1 = FS^1$ to all the rest $FS_i^1$ ($i$=2, ..., $n^1$) in Level 1.

**Step 3:** The same as Step 3 of Training Algorithm 1. ■

Fig. 8 illustrates the Training Algorithm 2 (for DCFS with parameter sharing) for the design of the Level 1 fuzzy systems $FS_1^1 = \cdots = FS_{n^1}^1 = FS^1$ (comparing Fig. 8 with Fig. 5 to see the difference between Training Algorithms 1 and 2). The design of the fuzzy systems $FS_1^l = \cdots = FS_{n^l}^l = FS^l$ in other levels ($l$=2, ..., $L$) can be illustrated in a similar way as in Figs. 6 and 7. A remark about the Training Algorithm 2 is as follows.

**Remark 4 (low memory requirement and fast training):** The computational load of Training Algorithm 2 (for DCFS with parameter sharing) is approximately $O(N\sum_{l=1}^{L} n^l + q^m L)$, where $O(N\sum_{l=1}^{L} n^l)$ accounts for the one-pass of the data (19) in the computation (13)-(15) for the data-covered $c^{j_1\cdots j_m}$'s, and $O(q^m L)$ accounts for the computation of the parameters $c^{j_1\cdots j_m}$ of the $L$ fuzzy systems $FS^l$ ($l$=2, ..., $L$) in Steps 2.5-2.7 (illustrated in Fig. 4). Comparing with the computational load $O(N+q^m)\sum_{l=1}^{L} n^l$ of Training Algorithm 1 (for general DCFS), we see that a computational reduction of $O(q^m)(\sum_{l=1}^{L} n^l - L)$ is achieved with Training Algorithm 2. For memory requirement (mainly for the storage of the parameters $c^{j_1\cdots j_m}$), it is reduced from $O(q^m\sum_{l=1}^{L} n^l)$ (for general DCFS) to $O(q^m L)$ (for DCFS with parameter sharing), a reduction of $O(q^m)(\sum_{l=1}^{L} n^l - L)$, which is in the same order as the computational reduction. For the DCFS in Fig. 3, for example, we have $L = 6$ and $\sum_{l=1}^{L} n^l = 21$, so the memory requirement is reduced from $21O(q^m)$ to $6O(q^m)$, a reduction of $\frac{(21-6)}{21} = 71.4\%$.

In Training Algorithms 1 and 2, the $N$ input-output data pairs (9) are processed in a batch fashion. In many real-world situations, the data are collected in real-time, so it is interesting to design the DCFS in a recursive way with new data coming in real-time. This gives Task 2 below.

**Task 2 (on-line training):** Let the input-output data pairs

$$[x_1^0(k), x_2^0(k), \ldots, x_n^0(k); y^0(k)], \quad k = 1, 2, 3, \ldots \tag{20}$$

be given on-line with $k$ being the time index. Suppose a DCFS has already be designed up to $k-1$, and we denote this DCFS as DCFS($k-1$). Let the initial DCFS(0) be designed either by Training Algorithm 1 for general DCFS or by Training Algorithm 2 for DCFS with parameter sharing, and our task is to update the DCFS($k-1$) based on the new data pair $[x_1^0(k), x_2^0(k), \ldots, x_n^0(k); y^0(k)]$ to get DCFS($k$). ■

We now show how to do Task 2 for the general DCFS and the DCFS with parameter sharing in Training Algorithms 3 and 4 below, respectively.

**Training Algorithm 3 (on-line training for general DCFS):** For Task 2, let the initial DCFS(0) be a general DCFS (Definition 1) designed by Training Algorithm 1 and $c^{j_1\cdots j_m}(k-1)$ be the parameters of the fuzzy systems $FS_i^l$ in the form of (7) in DCFS($k-1$). Then, the parameters $c^{j_1\cdots j_m}(k)$ of the $FS_i^l$ in DCFS($k$) are updated through the following steps:

**Step 1:** Update the $c^{j_1\cdots j_m}(k)$ of the Level 1 fuzzy systems $FS_i^1(x_i^0,\dots,x_{m+i-1}^0)$ in the form of (7) from $c^{j_1\cdots j_m}(k-1)$ using the new data pair $[x_i^0(k),\dots,x_{m-i+1}^0(k);y^0(k)]$. Specifically, determine

$$\begin{aligned} j_1^* &= \arg\max_{j\in\{1,2,\dots,q\}}\left(A^j\left(x_i^0(k)\right)\right), \\ &\vdots \\ j_m^* &= \arg\max_{j\in\{1,2,\dots,q\}}\left(A^j\left(x_{m+i-1}^0(k)\right)\right). \end{aligned} \tag{21}$$

and update

$$\begin{aligned} c^{j_1^*,\cdots,j_m^*}(k) &= \alpha\, A^{j_1^*}\left(x_i^0(k)\right)\cdots A^{j_m^*}\left(x_{m+i-1}^0(k)\right)\, y^0(k) + \\ &\left[1-\alpha\, A^{j_1^*}\left(x_i^0(k)\right)\cdots A^{j_m^*}\left(x_{m+i-1}^0(k)\right)\right] c^{j_1^*,\cdots,j_m^*}(k-1), \end{aligned} \tag{22}$$

where $\alpha\in[0,1]$ is a weighting factor. For other cells $(j_1,\cdots,j_m)$ $(j_1,\cdots,j_m=1,2,\dots,q)$ not equal to $(j_1^*,\cdots,j_m^*)$, set

$$c^{j_1,\cdots,j_m}(k)=c^{j_1,\cdots,j_m}(k-1) \tag{23}$$

**Step 2:** Pass the new data $x_1^0(k),x_2^0(k),\dots,x_n^0(k)$ through Level 1 to get the new input-output data pair $\left[x_1^1(k),x_2^1(k),\dots,x_{n^1}^1(k);y^0(k)\right]$ for the Level 2 fuzzy systems $FS_i^2$. Update the $c^{j_1\cdots j_m}(k)$ of the Level 2 fuzzy systems $FS_i^2(x_i^1,\dots,x_{m+i-1}^1)$ in the form of (7) from $c^{j_1\cdots j_m}(k-1)$ using the same (21)-(23) as in Step 1 with $x_i^0(k),\dots,x_{m-i+1}^0(k)$ replaced by $x_i^1(k),\dots,x_{m-i+1}^1(k)$, respectively.

**Step 3:** Repeat Step 2 for Levels 3, 4, …, until the top Level L.

■

**Training Algorithm 4 (on-line training for DCFS with parameter sharing):** For Task 2, let the initial DCFS(0) be a parameter sharing DCFS (Definition 2) designed by Training Algorithm 2 and $c^{j_1\cdots j_m}(k-1)$ be the parameters of the fuzzy systems $FS^l$ in the form of (7) in DCFS($k-1$). Then, the parameters $c^{j_1\cdots j_m}(k)$ of the $FS^l$ in DCFS($k$) are updated through the following steps:

**Step 1:** Update the $c^{j_1\cdots j_m}(k)$ of the first Level 1 fuzzy system $FS_1^1=FS^1$ in the form of (7) with the following $n^1$ input-output data pairs:

$$[x_i^0(k),\dots,x_{m+i-1}^0(k);y^0(k)],\quad i=1,\dots,n^1. \tag{24}$$

Specifically, for each of the $n^1$ input-output data pairs of (24), update the $c^{j_1^*,\cdots,j_m^*}(k)$ using (21)-(22); for the rest $c^{j_1\cdots j_m}(k)$, keep them the same using (23). Then, copy this $FS_1^1=FS^1$ to all the rest $FS_i^1$ ($i$=2, …, $n^1$) in Level 1.

**Steps 2 and 3:** The same as Steps 2 and 3 of Training Algorithm 3, with the Step 1 there replaced by the Step 1 here.

■

Two remarks about the Training Algorithms 3 and 4 are now in order.

**Remark 5 (physical meaning and very low computational cost of the updating law):** From (21)-(23) we see that for each new input-output data pair, only the $c^{j_1^*,\cdots,j_m^*}$ that covers the new input data point $\left(x_i^0(k),\dots,x_{m+i-1}^0(k)\right)$ is updated, and the rest $c^{j_1\cdots j_m}$'s remain unchanged. The new $c^{j_1^*,\cdots,j_m^*}(k)$ (22) is a weighted average of the output $y^0(k)$ and the previous $c^{j_1^*,\cdots,j_m^*}(k-1)$ with weight $\alpha\, A^{j_1^*}\left(x_i^0(k)\right)\cdots A^{j_m^*}\left(x_{m+i-1}^0(k)\right)$ which is a measure of the degree that the input data point $\left(x_i^0(k),\dots,x_{m+i-1}^0(k)\right)$ belongs to the cell $(j_1^*,\cdots,j_m^*)$. Since only one parameter $c^{j_1^*,\cdots,j_m^*}(k)$ needs to be updated and the computational load of the updating law (21)-(22) is very low, Training Algorithms 3 and 4 may be implemented on the simple devices such as a mobile phone. Therefore, an initial DCFS designed with the off-line Training Algorithms 1 or 2 may be downloaded to the customer's simple device such as a mobile phone, then the customer may updated the DCFS using the on-line Training Algorithms 3 or 4 with the new data that the customer observes in real life.

**Remark 6 (easy error correction):** As we discussed in the Introduction that a main problem of the black-box DCNN is that if something goes wrong, we don't know which parts of the DCNN should be corrected so that the same mistake will not take place again. The DCFS designed with the Training Algorithms 3 or 4 can easily solve this problem of the DCNN. Specifically, if a DCFS makes a mistake at an input point, say the point $(x_1^{0*},\dots,x_7^{0*})$ to the DCFS in Fig. 3, then we can update the $c^{ij}$'s of the DCFS in Fig. 3 that are responsible for the mistake, using Training Algorithms 3 or 4 with the correct input-output data.

We now apply the DCFS designed with Training Algorithms 1- 4 to the time-series prediction problems in the next section.

## IV. Application to Hang Seng Index Prediction

Before we apply the DCFS with the Training Algorithms to predict the real Hang Seng Index (HSI) of the Hong Kong stock market in Examples 2, 4 and 5 below, we first try it for a synthetic chaotic time-series in Examples 1 and 3 to get some feeling about the performance of the method. Specifically, in Examples 1 and 2 below, we test the Training Algorithms 1 and 2 for predicting the Mackey-Glass chaotic time-series and the real Hang Seng Index, respectively. In Examples 3 and 4, we apply the on-line Training Algorithms 3 and 4 to predict the Mackey-Glass chaotic time-series and the real Hang Seng Index, respectively. In Example 5, we add more related stocks to the inputs to get better performance.

**Example 1:** Consider the Mackey-Glass chaotic time-series generated by the differential equation:

$$\frac{dx(t)}{dt}=\frac{0.2x(t-\tau)}{1+x^{10}(t-\tau)}-0.1x(t) \tag{25}$$

with $\tau=50$. Let $r(t)$ be the return (relative change) of the chaotic time-series $x(t)$ plus a white Gaussian noise $n(t)$, i.e.,

$$r(t)=\frac{x(t)-x(t-1)}{x(t-1)}+n(t), \tag{26}$$

and we generate a synthetic chaotic plus random time-series $y(t)$ whose return is $r(t)$, i.e.,

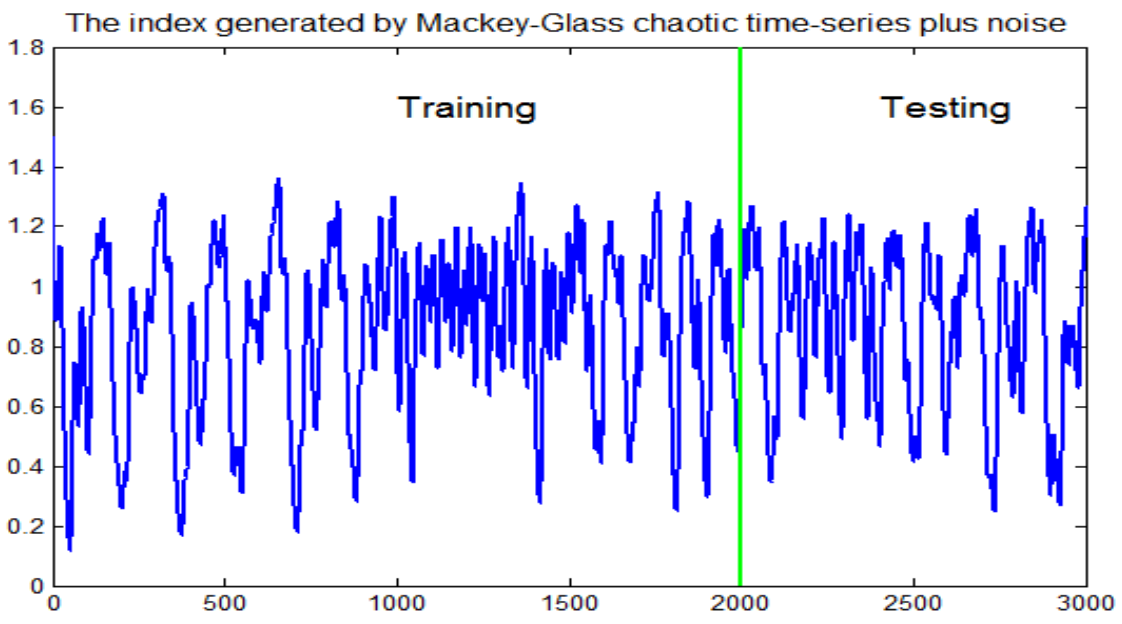

Fig. 9: The index generated by the Mackey-Glass chaotic time-series plus noise, where the first 2000 points are training data and the last 1000 points are testing data.

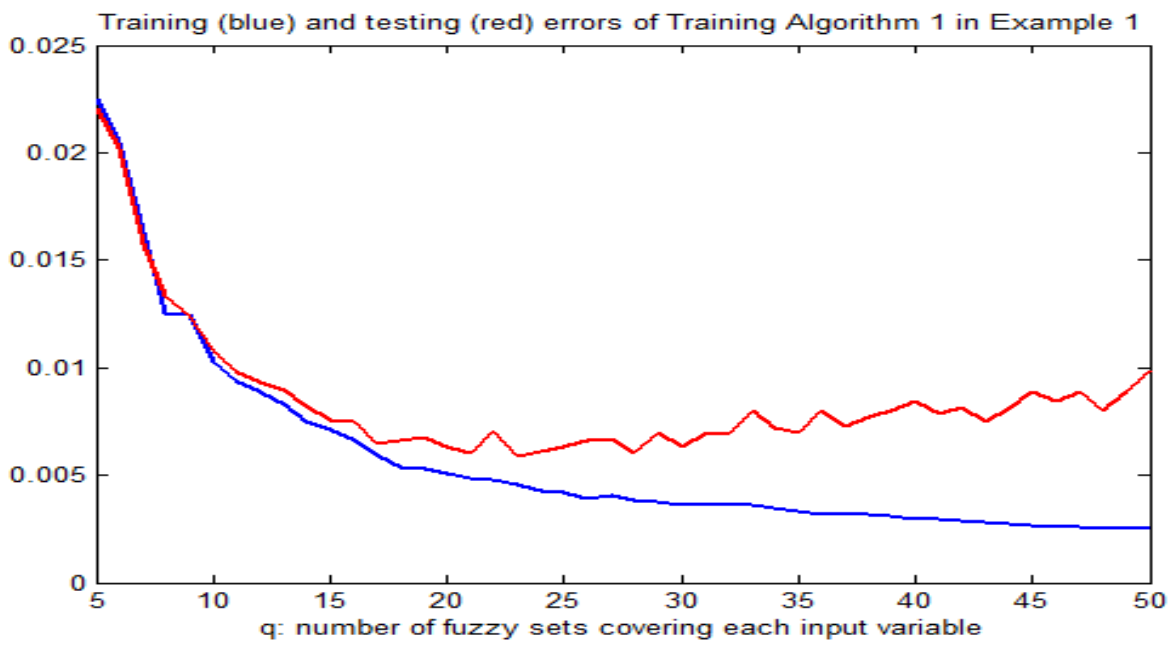

Fig. 10: The training (blue) and testing (red) errors of Training Algorithm 1 (for general DCFS) as function of $q$ for predicting the Mackey-Glass chaotic time-series.

$$y(t) = y(t-1)[1 + r(t)]. \tag{27}$$

We view the $y(t)$ $(t = 0,1,2,\dots)$ of (27) as the daily closing prices of a stock index. Fig. 9 plots a realization of 3000 points of $y(t)$.

We now use the DCFS of Fig. 1 with the Training Algorithms 1 and 2 to predict the return sequence $r(t)$ of (26). Specifically, let the $n$ past returns up to day $t-1$: $r(t-1)$, $r(t-2),\dots,r(t-n)$ be the inputs $x_1^0, x_2^0, \dots, x_n^0$ to the DCFS and the output $x^L$ of the DCFS be the prediction of the return at day $t$, i.e.,

$$\hat{r}(t) = \text{DCFS}\big(r(t-1), r(t-2),\dots,r(t-n)\big), \tag{28}$$

where $\hat{r}(t)$ is the prediction of $r(t)$. In this case, the input-output data pairs of (9) become

$$[r(k-1), r(k-2),\dots,r(k-n); r(k)], \tag{29}$$

where $k = t-1, t-2, \dots, t-N+n$ (the current day is $t-1$); that is, at day $t-1$, $N$ past returns $r(t-1), r(t-2),\dots,$ $r(t-N)$ constitute the $N-n$ input-output pairs in the form of (29).

With $n = 11$, $m = 3$ and one-variable-at-a-time moving scheme, a 5-level DCFS is established, where Level 1 to Level 5 have 9, 7, 5, 3 and 1 fuzzy systems $FS_i^l$, respectively. Using the first 2000 points in Fig. 9 as training data and the last 1000 points as testing data, we simulate the Training Algorithms 1 and 2 for different values of $q$ ($q$ is the number of fuzzy sets covering each input variable, see Fig. 2). Figs. 10 and 11 plot the training (blue) and testing (red) errors of Training Algorithm 1 (for general DCFS) and Training Algorithm 2 (for DCFS with parameter sharing) as function of $q$, respectively. We see from Figs. 10 and 11 that for both general DCFS (with Training Algorithm 1) and DCFS with parameter sharing (with Training Algorithm 2), the training errors (blue curves) keep decreasing as more fuzzy sets are used to cover the input variables. For the testing errors (red curves in Figs. 10 and 11), we see that they are first decreasing as $q$ increases, but then begin to increase as the DCFS models overfit the data.

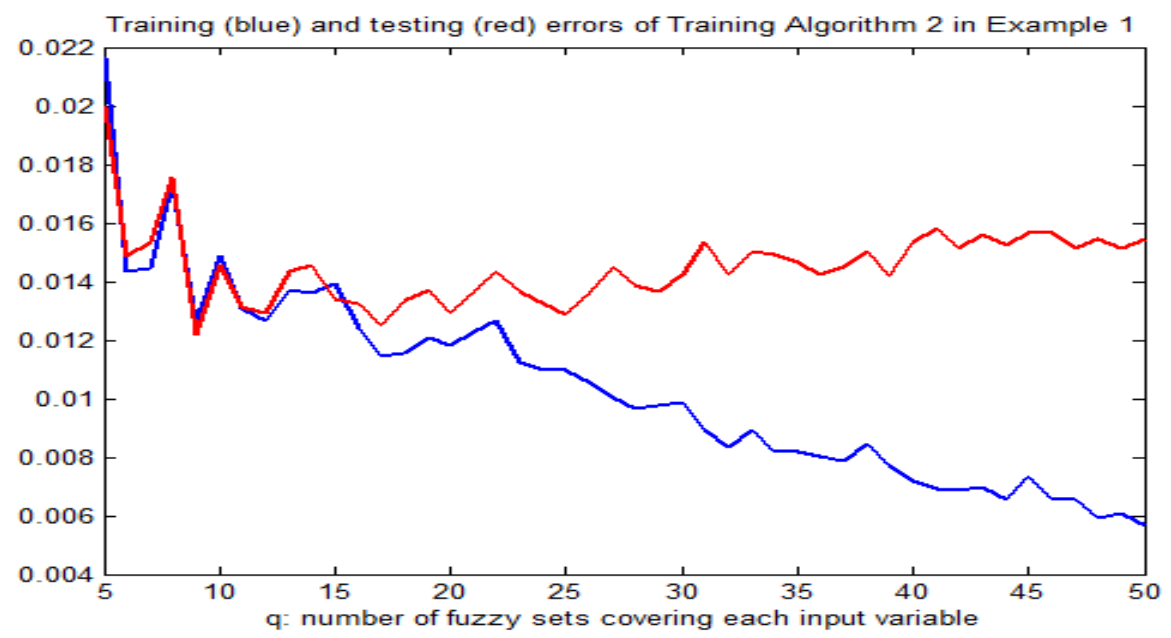

Fig. 11: The training (blue) and testing (red) errors of Training Algorithm 2 (for DCFS with parameter sharing) as function of $q$ for predicting the Mackey-Glass chaotic time-series.

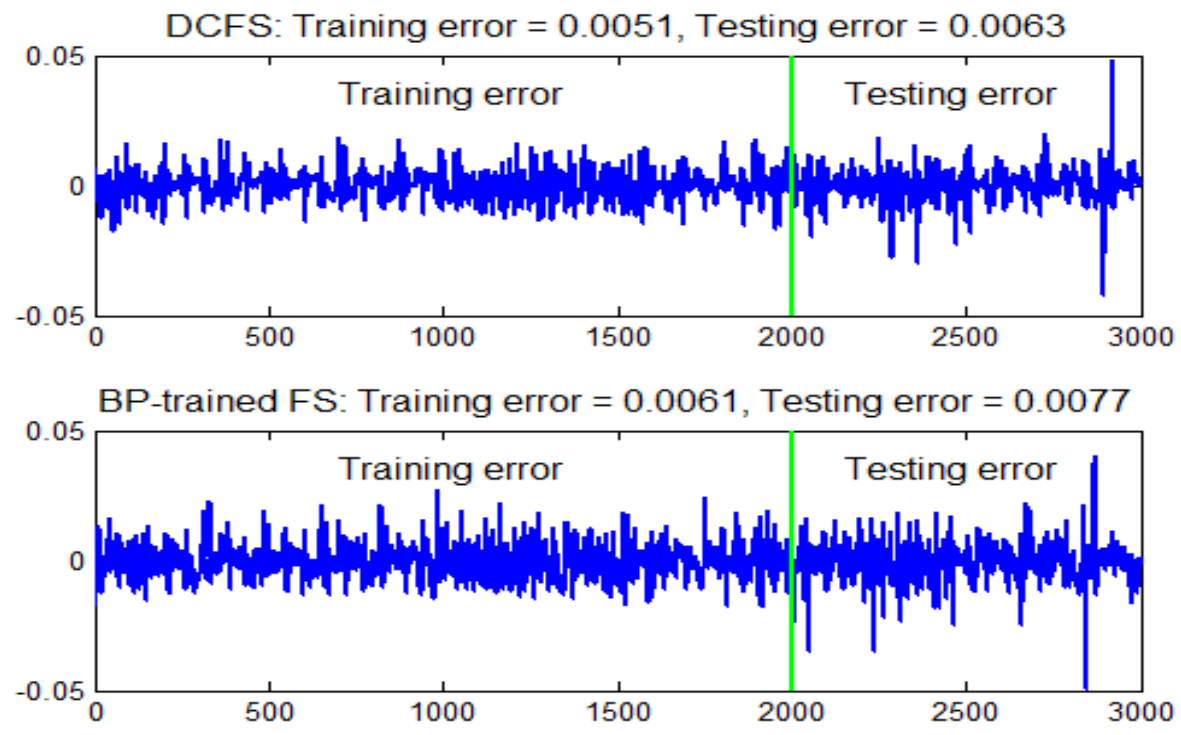

Fig. 12: The training (first 2000 points) and testing (last 1000 points) errors of the DCFS predictor (top) and the BP trained fuzzy model (bottom) for the 3000 Mackey-Glass chaotic time-series data in Fig. 9.

We also compare the DCFS predictor with the fuzzy model trained by the back-propagation (BP) algorithm [60, 61]. For models with similar complexity (roughly the same number of free parameters), the training and testing errors of the DCFS predictor are about 20% less than that of the BP trained fuzzy model, with a training speed at least ten times faster. Specifically, for $q = 20$ (roughly $q^m \sum_{l=1}^{L} n^l = 20^3 25 =$ 200,000 free parameters) and the 3000 chaotic time-series data in Fig. 9, the training (first 2000 points) and testing (last 1000 points) errors of the DCFS predictor and the BP trained fuzzy model are plotted in Fig. 12, where the mean square training and testing errors of the DCFS predictor are 0.0051 and 0.0063, respectively, and those of the BP trained fuzzy model are 0.0061 and 0.0077, respectively. ■

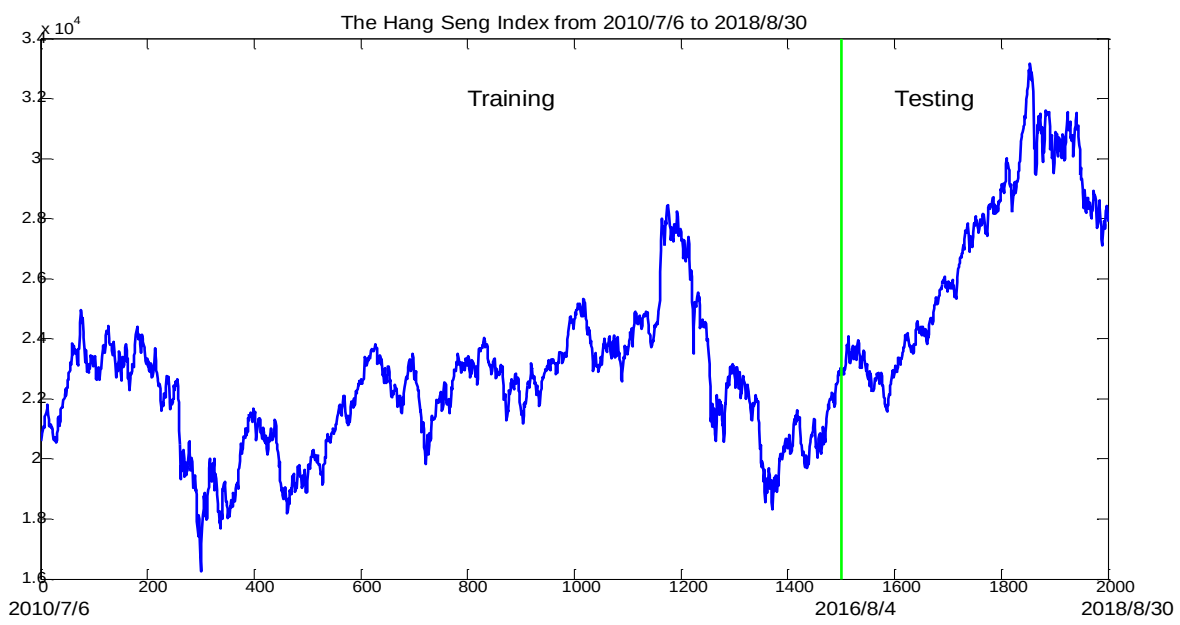

Fig. 13: The Hang Seng Index daily closing from 2010/7/6 to 2018/8/30, where the first 1500 points are training data and the last 500 points are testing data.

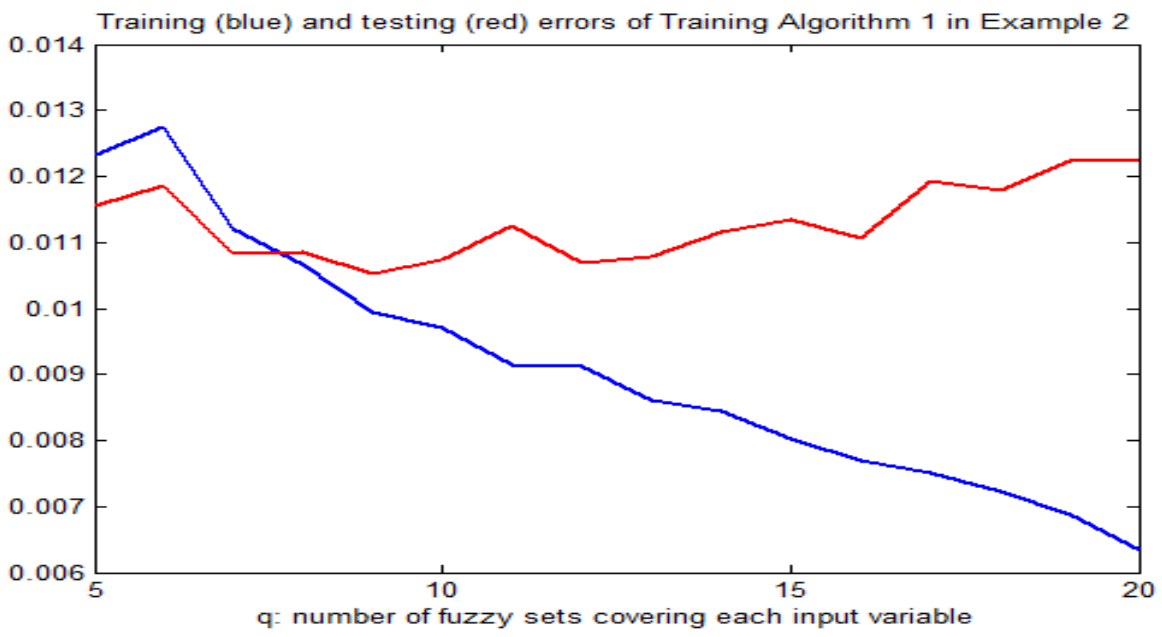

Fig. 14: The training (blue) and testing (red) errors of Training Algorithm 1 as function of $q$ for predicting the Hang Seng Index.

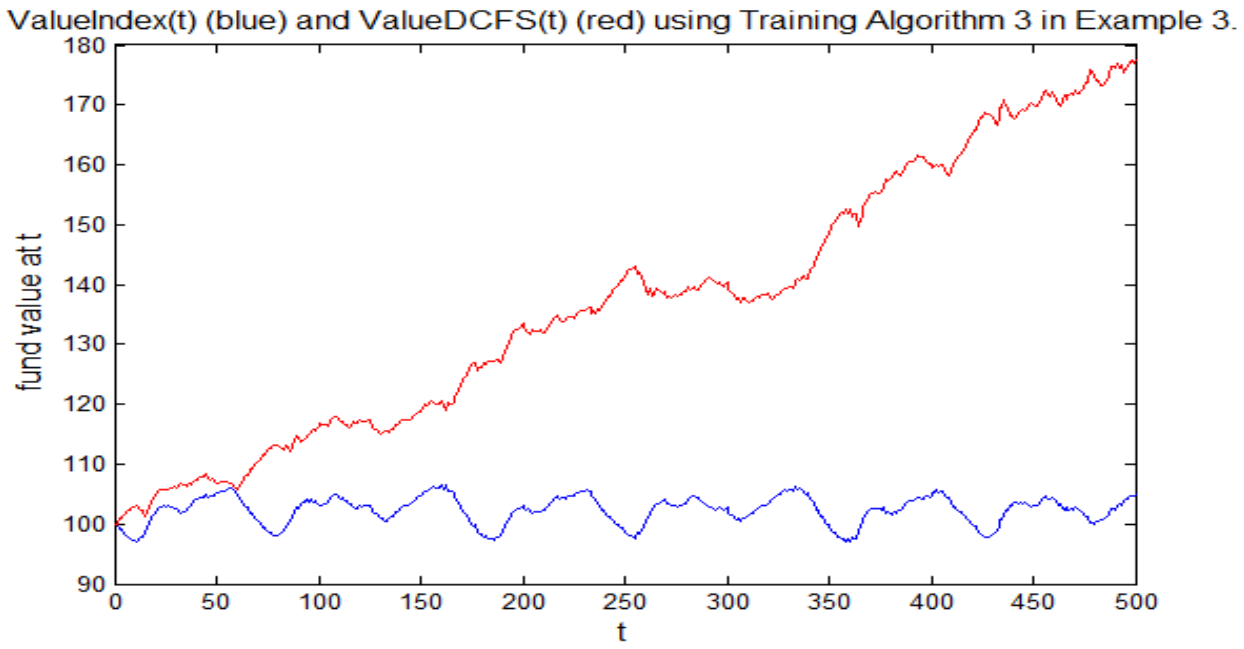

Fig. 15: ValueIndex$(t)$ (blue line) and ValueDCFS$(t)$ (red line) with Training Algorithm 3 in Example 3.

**Example 2:** The same as Example 1 except that the index generated by the Mackey-Glass chaotic time-series plus noise (Fig. 9) is replaced by the real Hang Seng Index (Fig. 13). Specifically, considering the Hang Seng Index daily closing from 2010/7/6 to 2018/8/30 (2000 data points) plotted in Fig. 13, we use the first 1500 points (from 2010/7/6 to 2016/8/4) as training data and the last 500 points (from 2016/8/5 to 2018/8/30) as testing data. Fig. 14 plots the training (blue) and testing (red) errors of Training Algorithm 1 (for general DCFS) as function of $q$. Comparing Fig. 14 with Fig. 10, we see that the real Hang Seng Index is much more difficult to predict than the index generated by the Mackey-Glass chaotic time-series. ■

Since we view the $y(t)$ of (27) as the daily closing prices of a stock index, the value of an Index Fund with the $y(t)$ at day $t$, denoted as ValueIndex$(t)$, is updated daily according to

$$\text{ValueIndex}(t) = \text{ValueIndex}(t-1)[1 + r(t)], \quad (30)$$

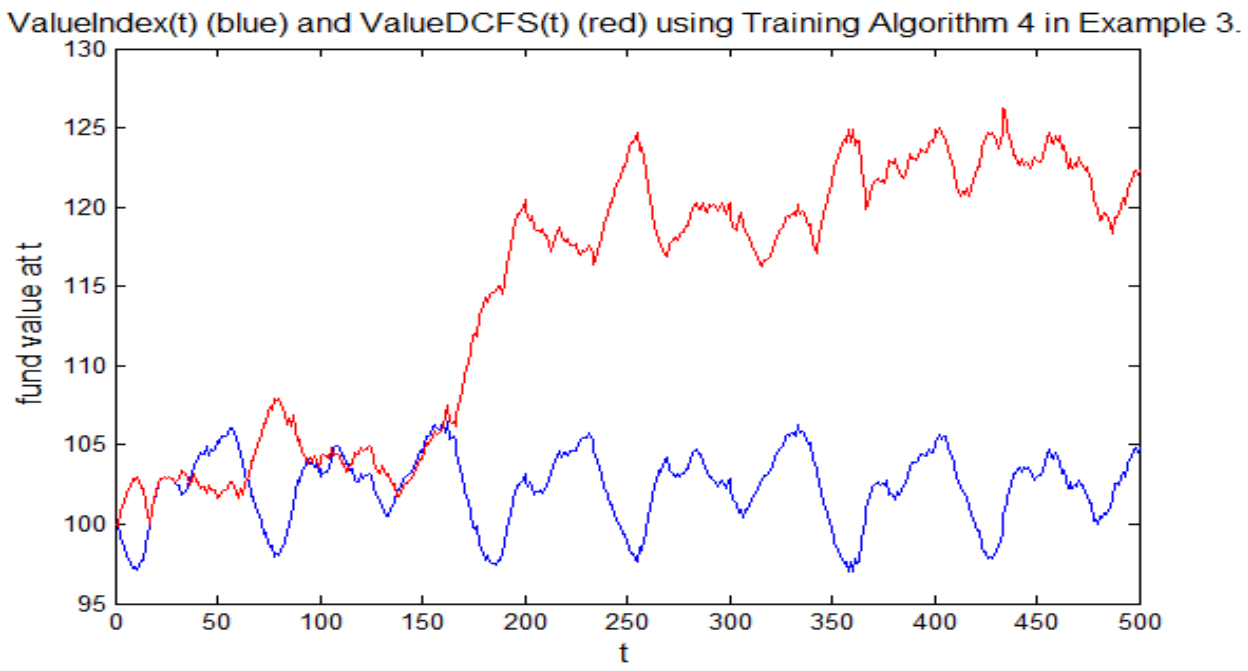

Fig. 16: ValueIndex$(t)$ (blue line) and ValueDCFS$(t)$ (red line) with Training Algorithm 4 in Example 3.

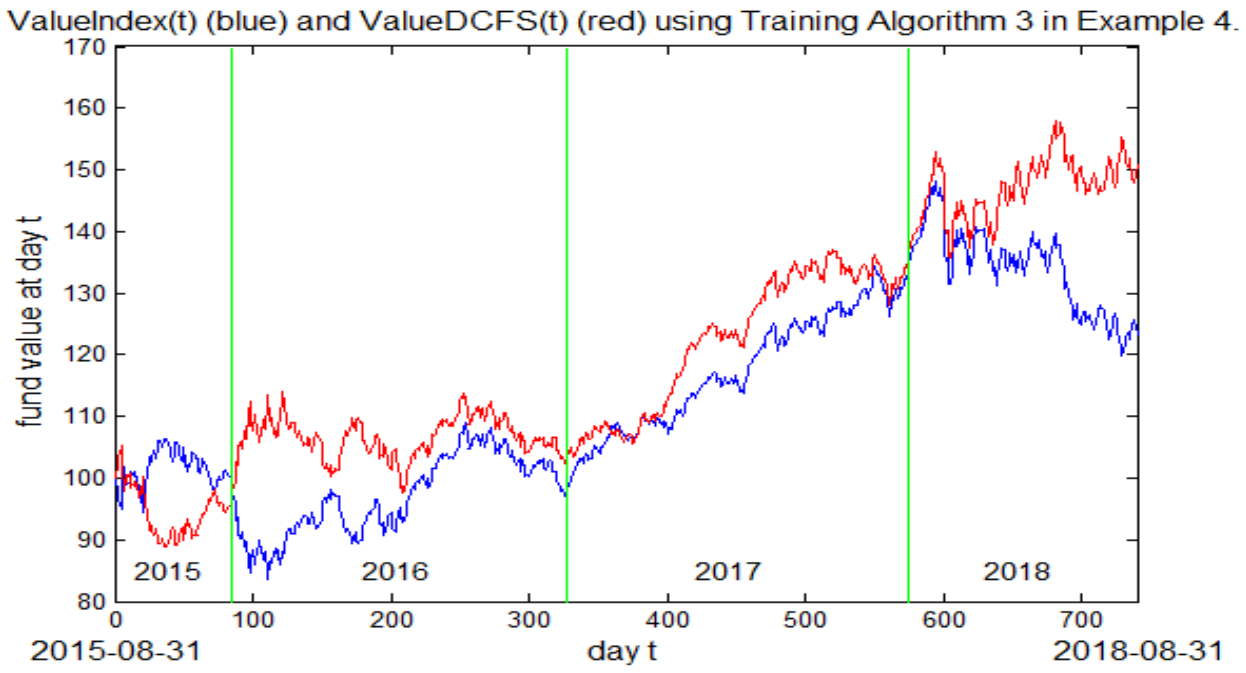

Fig. 17: ValueIndex$(t)$ (blue line) and ValueDCFS$(t)$ (red line) with Training Algorithm 3 in Example 4.

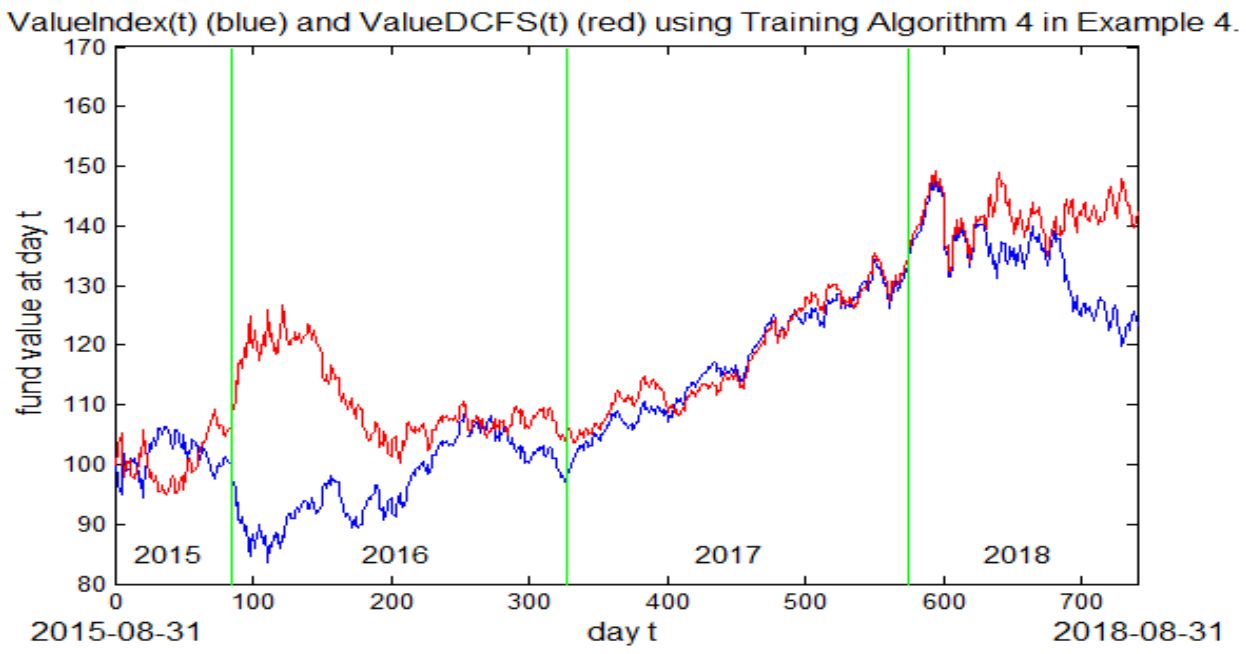

Fig. 18: ValueIndex$(t)$ (blue line) and ValueDCFS$(t)$ (red line) with Training Algorithm 4 in Example 4.

where $t = 1,2,\ldots$, and we assume the initial investment ValueIndex$(0) = 100$.

We now propose a trading strategy based on the DCFS prediction $\hat{r}(t)$ of (28).

**Trading Strategy based on DCFS Prediction:** At day $t-1$ (the most recent return available is $r(t-1)$), use the Training Algorithms 3 or 4 in Section III to design a DCFS of (28) to get the prediction $\hat{r}(t)$ of the return at day $t$. If $\hat{r}(t) > 0$, meaning that we predict the index $y(t)$ will go up in day $t$, then long (buy) the index at day $t-1$; if $\hat{r}(t) < 0$, meaning that we predict the index $y(t)$ will go down in day $t$, then short (sell) the index at day $t-1$.

The value of the fund with this trading strategy at day $t$, denoted as ValueDCFS$(t)$, is

$$\text{ValueDCFS}(t) = \text{ValueDCFS}(t-1)[1 + I(t)|r(t)|], \quad (31)$$

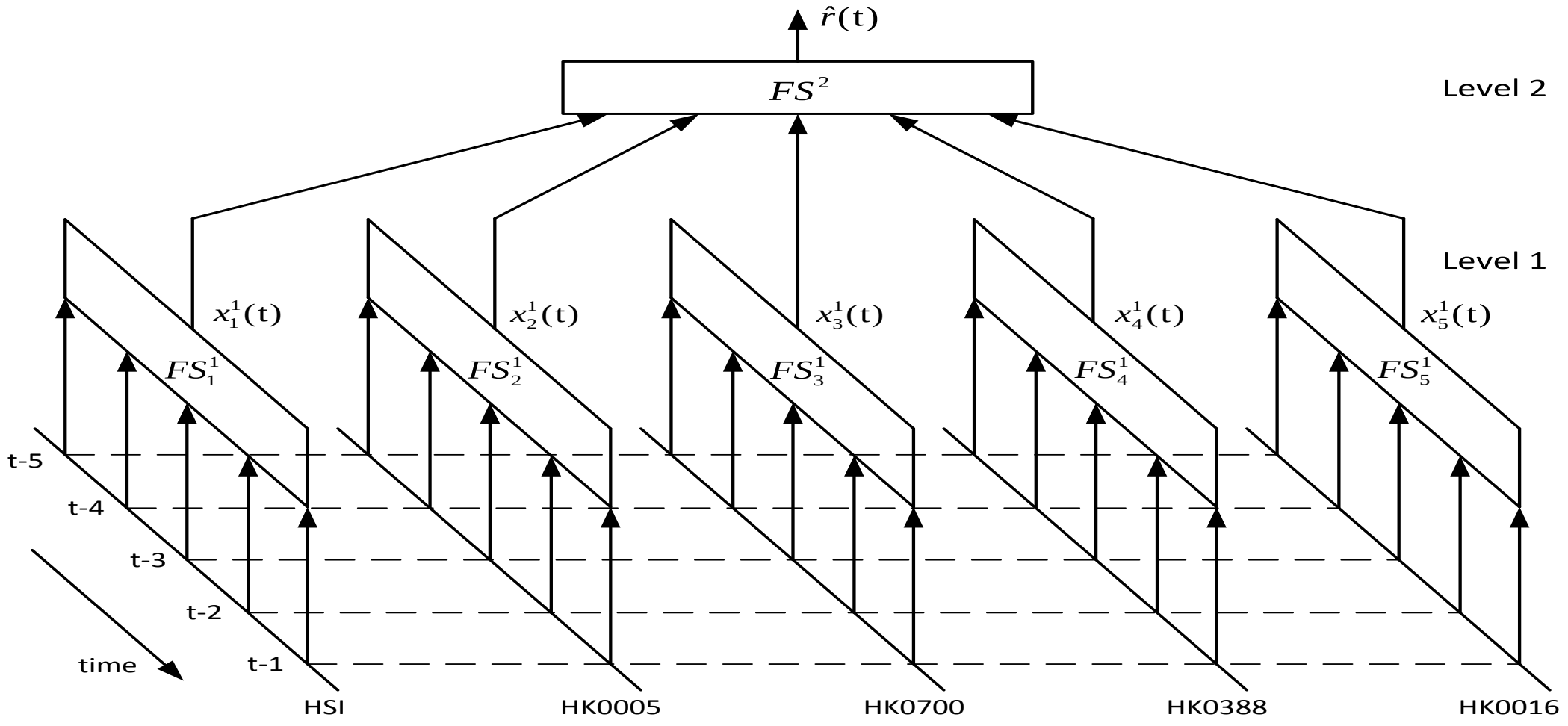

Fig. 19: The two-level DCFS in Example 5, where the HSI returns and the returns of other four major stocks are used to construct the five fuzzy systems in the first-level whose outputs are combined by the second-level fuzzy system to form the final prediction of the HSI return.

where the initial investment $\text{ValueDCFS}(0) = 100$ and

$$I(t) = \begin{cases} 1 & \text{if } \text{sign}(\hat{r}(t)) = \text{sign}(r(t)) \\ -1 & \text{if } \text{sign}(\hat{r}(t)) \neq \text{sign}(r(t)). \end{cases} \quad (32)$$

The meaning of (31) and (32) is that: if $\text{sign}(\hat{r}(t)) = \text{sign}(r(t))$ which means we made the right prediction at day $t-1$, then our fund value will increase by $|r(t)|$ no matter the index $y(t)$ goes up or down at day $t$; on the other hand, if $\text{sign}(\hat{r}(t)) \neq \text{sign}(r(t))$ which means we made a wrong prediction at day $t-1$, then our fund will decrease in value by $|r(t)|$ no matter the index $y(t)$ goes up or down at day $t$.

We now test this trading strategy for the chaotic plus random index of Fig. 9 and the real Hang Seng Index of Fig. 13 in Examples 3 and 4 below, respectively.

**Example 3:** The same as in Example 1 except that Training Algorithms 3 or 4 is used to build the DCFS predictor (28). Figs. 15 and 16 plot the index fund value $\text{ValueIndex}(t)$ of (30) and the DCFS fund value $\text{ValueDCFS}(t)$ of (31) using Training Algorithms 3 and 4, respectively. We see from Figs. 15 and 16 that: 1) in both cases the DCFS fund performs much better than the index fund, and 2) the general DCFS with Training Algorithm 3 is better than the parameter sharing DCFS with Training Algorithm 4. ■

**Example 4:** The same as Example 3 except that the index generated by the Mackey-Glass chaotic time-series plus noise (Fig. 9) is replaced by the real Hang Seng Index (Fig. 13). Figs. 17 and 18 plot the index fund value $\text{ValueIndex}(t)$ and the DCFS fund value $\text{ValueDCFS}(t)$ using Training Algorithms 3 and 4, respectively. We see from Figs. 17 and 18 that the DCFS fund performs slightly better than the index fund in both cases, and the general DCFS with Training Algorithm 3 is better than the parameter sharing DCFS with Training Algorithm 4. ■

To improve the performance of the DCFS fund, we notice that there may be other factors that influence the real Hang Seng Index returns $r(t)$, in addition to the past returns $r(t-1), r(t-2), \ldots, r(t-n)$ used in the DCFS model (28). For example, the returns of the major stocks at the current day $t-1$ may have a stronger influence on the tomorrow's HSI return $r(t)$ than the HSI returns $r(t-n)$ long in the past (when $n$ is large). Indeed, the long-past HSI returns $r(t-n)$ may introduce more noise than useful information for the prediction of tomorrow's HSI return $r(t)$. Hence, we add the returns of four major stocks in the Hong Kong stock market to the input space of the DCFS model and use only the five most recent returns as the inputs to the prediction model in the next example.

**Example 5**: We use the two-level DCFS in Fig. 19 as the prediction model for the HSI return $r(t)$, where the five fuzzy systems $FS_1^1, FS_2^1, \ldots, FS_5^1$ in Level-1 use the five past daily returns of HSI, HK0005 (HSBC Holdings plc), HK0700 (Tencent Holdings Limited), HK0388 (Hong Kong Exchanges and Clearing Limited) and HK0016 (Sun Hung Kai Properties Limited), respectively, as the inputs to produce five weak estimators $x_1^1(t), x_2^1(t), \ldots, x_5^1(t)$ for the HSI return $r(t)$, i.e.,

$$x_1^1(t) = FS_1^1(r_{HSI}(t-1), \ldots, r_{HSI}(t-5)) \quad (33)$$
$$x_2^1(t) = FS_2^1(r_{HK0005}(t-1), \ldots, r_{HK0005}(t-5)) \quad (34)$$
$$x_3^1(t) = FS_3^1(r_{HK0700}(t-1), \ldots, r_{HK0700}(t-5)) \quad (35)$$
$$x_4^1(t) = FS_4^1(r_{HK0388}(t-1), \ldots, r_{HK0388}(t-5)) \quad (36)$$
$$x_5^1(t) = FS_5^1(r_{HK0016}(t-1), \ldots, r_{HK0016}(t-5)) \quad (37)$$

where $r_{HSI}$, $r_{HK0005}$, $r_{HK0700,}$ $r_{HK0388}$ and $r_{HK0016}$ are the returns of the HSI and the stocks HK0005, HK0700, HK0388 and HK0016, respectively, then these five weak estimators are combined by the Level-2 fuzzy system $FS^2$ to produce the final prediction:

$$\hat{r}(t) = FS^2(x_1^1(t), x_2^1(t), \ldots, x_5^1(t)). \quad (38)$$

Fig. 20 plots the HSI fund value $\text{ValueIndex}(t)$ of (30) and the DCFS fund value $\text{ValueDCFS}(t)$ of (31) with this two-level DCFS predictor (33)-(38) using Training Algorithm 3 for the real HSI daily returns $r(t)$ over the three years from 31-Aug-2015 to 31-Aug-2018. Comparing the DCFS fund values $\text{ValueDCFS}(t)$ in Figs. 17 and 20 we see a clear performance improvement by adding these four stocks into the input space of the DCFS predictor. ■

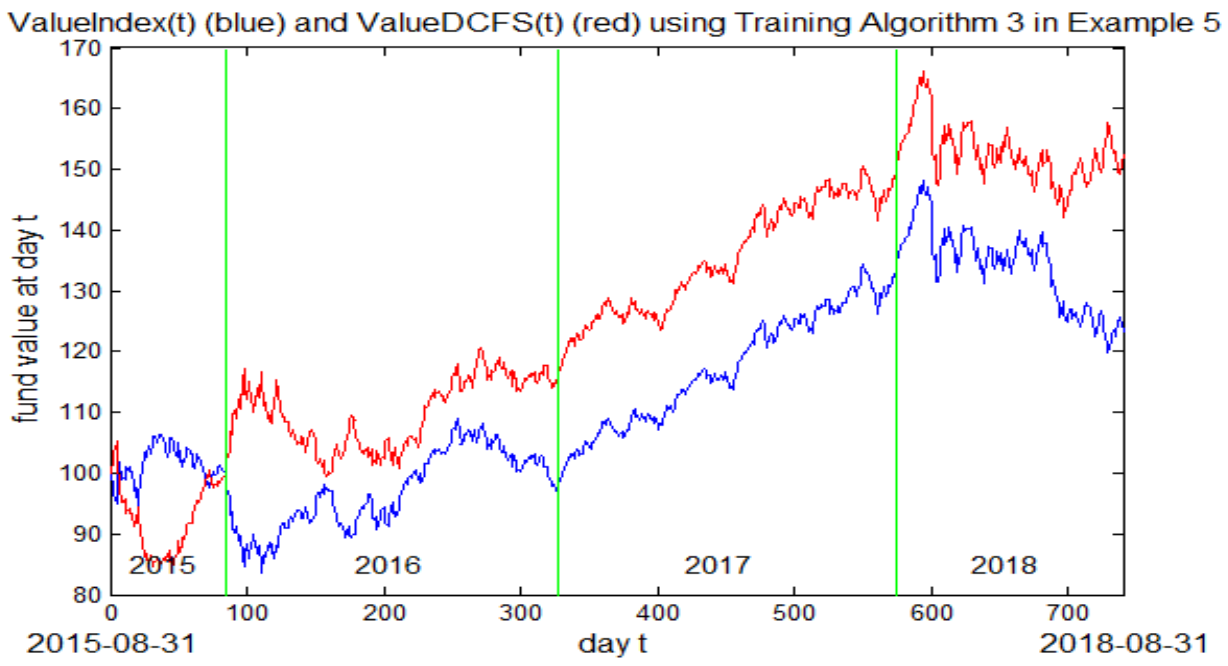

Fig. 20: ValueIndex$(t)$ (blue line) and ValueDCFS$(t)$ (red line) with Training Algorithm 3 in Example 5.

## V. Concluding Remarks

The deep convolutional fuzzy system (DCFS) models and the Training Algorithms proposed in this paper have the following advantages:

(i) It is fast: the data are used only once in the design of the fuzzy systems in the DCFS and no iterative training is needed;

(ii) It is highly interpretable: the fuzzy systems in different levels of the DCFS are weak estimators of the output variable that are constructed in a layer-by-layer, bottom-up fashion;

(iii) It is very flexible: the size of the moving window, the steps of each move, the number of fuzzy sets to cover the input variables and the number of layers can be easily adjusted for better performance;

(iv) It is easy to correct mistakes: because of the clear physical meaning of the parameters, it is easy to redesign the parameters with new data so that the same mistake will not happen again;

(v) It may be implemented on simple devices: due to simple computation and low memory requirement, the on-line training algorithms may be run on the user-end simple devices such as a mobile phone;

(vi) It supports on-line learning: users can continuously update the DCFS models with their own new data on their own simple devices, so that a wide variety of user-specific intelligent systems would be created;

(vii) It provides a natural structure for parallel computing: all the fuzzy systems in the same level can be trained in parallel, making the fast training algorithms even faster through parallel computing; and,

(viii) It is suitable for high-dimensional problems.

## Acknowledgements

The author wishes to thank the reviewers for their very insightful comments that helped to improve the paper.

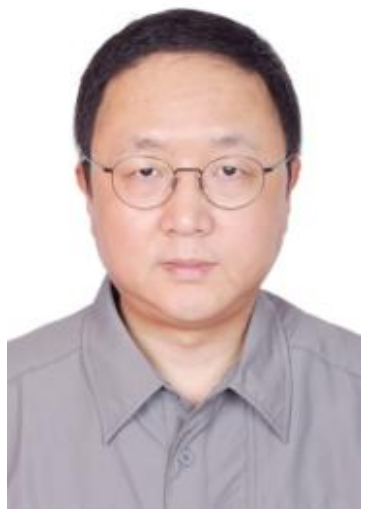

**Li-Xin Wang** received the Ph.D. degree from the Department of Electrical Engineering, University of Southern California (USC), Los Angeles, CA, USA, in 1992.

From 1992 to 1993, he was a Postdoctoral Fellow with the Department of Electrical Engineering and Computer Science, University of California at Berkeley. From 1993 to 2007, he was on the faculty of the Department of Electronic and Computer Engineering, The Hong Kong University of Science and Technology (HKUST). In 2007, he resigned from his tenured position at HKUST to become an independent researcher and investor in the stock and real estate markets in Hong Kong and China. In Fall 2013, he returned to academia and is now a professor with the University of Chinese Academy of Sciences, Beijing, China. His research interests are stock dynamics, market microstructure, trading strategies, fuzzy systems, and opinion networks.

## THE MATLAB CODE OF TRAINING ALGORITHM 1 FOR THE DCFS IN EXAMPLE 1 (11 INPUTS 5 LEVELS)

```
% The following 13 lines generate data file xx for Example 1 (predict
% the Mackey-Glass time-series plus noise). You may use other data
% file xx and start from load xx.
for k=1:50
   p0(k)=0.04*(k-1);
end;
for k=51:3100
   p0(k)=0.9*p0(k-1)+0.2*p0(k-50)/(1+p0(k-50)^10);
   r0(k-50)=log(p0(k)/p0(k-1))+0.0001*randn;
end;
for i=1:3000
   for j=1:12
      xx(i,j)=r0(i+j);
   end;
end;
save xx;

load xx; % xx is N×12 matrix containing all the data, the first
           % 11 columns are inputs and the last column is output.
[N,~]=size(xx);
ntrain=round(N*2/3); % The first 2/3 are training data, and the
                     % remaining 1/3 are testing data.
mm=20; % mm is the q in Fig. 2: the number of fuzzy sets covering
           % the input variables.
for i=1:ntrain
   for j=1:3 % window size=3 with one-variable-at-a-time moving.
      x11(i,j)=xx(i,j);
      x12(i,j)=xx(i,j+1);
      x13(i,j)=xx(i,j+2);
      x14(i,j)=xx(i,j+3);
      x15(i,j)=xx(i,j+4);
      x16(i,j)=xx(i,j+5);
      x17(i,j)=xx(i,j+6);
      x18(i,j)=xx(i,j+7);
      x19(i,j)=xx(i,j+8);
   end; % x11,…,x19 are inputs to the 9 fuzzy systems in Level 1.
   y(i,1)=xx(i,12); % y is the desired output.
end;
```

% The following 9 lines train the 9 fuzzy systems in Level 1, where
% wmdeepzb trains the fuzzy system to get zb (the $c^{j_1 \cdots j_m}$ in FS (7))
% and rg (the endpoints in (11)).

```
[zb11 rg11]=wmdeepzb(mm,x11,y);
[zb12 rg12]=wmdeepzb(mm,x12,y);
[zb13 rg13]=wmdeepzb(mm,x13,y);
[zb14 rg14]=wmdeepzb(mm,x14,y);
[zb15 rg15]=wmdeepzb(mm,x15,y);
[zb16 rg16]=wmdeepzb(mm,x16,y);
[zb17 rg17]=wmdeepzb(mm,x17,y);
[zb18 rg18]=wmdeepzb(mm,x18,y);
[zb19 rg19]=wmdeepzb(mm,x19,y);
'Level 1 training done '

% The following computes x21,…,x27 which are inputs to the 7 fuzzy
% systems in Level 2, where wmdeepyy computes the output of the
% fuzzy system with parameters zb and rg.
x21(1:ntrain,1)=wmdeepyy(mm,zb11,rg11,x11);
x21(1:ntrain,2)=wmdeepyy(mm,zb12,rg12,x12);
x21(1:ntrain,3)=wmdeepyy(mm,zb13,rg13,x13);
x22(1:ntrain,1)=x21(:,2);
x22(1:ntrain,2)=x21(:,3);
x22(1:ntrain,3)=wmdeepyy(mm,zb14,rg14,x14);
x23(1:ntrain,1)=x21(:,3);
x23(1:ntrain,2)=x22(:,3);
x23(1:ntrain,3)=wmdeepyy(mm,zb15,rg15,x15);
x24(1:ntrain,1)=x22(:,3);
x24(1:ntrain,2)=x23(:,3);
x24(1:ntrain,3)=wmdeepyy(mm,zb16,rg16,x16);
x25(1:ntrain,1)=x23(:,3);
x25(1:ntrain,2)=x24(:,3);
x25(1:ntrain,3)=wmdeepyy(mm,zb17,rg17,x17);
x26(1:ntrain,1)=x24(:,3);
x26(1:ntrain,2)=x25(:,3);
x26(1:ntrain,3)=wmdeepyy(mm,zb18,rg18,x18);
x27(1:ntrain,1)=x25(:,3);
x27(1:ntrain,2)=x26(:,3);
x27(1:ntrain,3)=wmdeepyy(mm,zb19,rg19,x19);
% The following 7 lines train the 7 fuzzy systems in Level 2.
[zb21 rg21]=wmdeepzb(mm,x21,y);
[zb22 rg22]=wmdeepzb(mm,x22,y);
[zb23 rg23]=wmdeepzb(mm,x23,y);
[zb24 rg24]=wmdeepzb(mm,x24,y);
[zb25 rg25]=wmdeepzb(mm,x25,y);
[zb26 rg26]=wmdeepzb(mm,x26,y);
[zb27 rg27]=wmdeepzb(mm,x27,y);
'Level 2 training done '

% x31,…,x35 are inputs to the 5 fuzzy systems in Level 3.
x31(1:ntrain,1)=wmdeepyy(mm,zb21,rg21,x21);
x31(1:ntrain,2)=wmdeepyy(mm,zb22,rg22,x22);
x31(1:ntrain,3)=wmdeepyy(mm,zb23,rg23,x23);
x32(1:ntrain,1)=x31(:,2);
x32(1:ntrain,2)=x31(:,3);
x32(1:ntrain,3)=wmdeepyy(mm,zb24,rg24,x24);
x33(1:ntrain,1)=x31(:,3);
x33(1:ntrain,2)=x32(:,3);
x33(1:ntrain,3)=wmdeepyy(mm,zb25,rg25,x25);
x34(1:ntrain,1)=x32(:,3);
x34(1:ntrain,2)=x33(:,3);
x34(1:ntrain,3)=wmdeepyy(mm,zb26,rg26,x26);
x35(1:ntrain,1)=x33(:,3);
x35(1:ntrain,2)=x34(:,3);
x35(1:ntrain,3)=wmdeepyy(mm,zb27,rg27,x27);
% The following 5 lines train the 5 fuzzy systems in Level 3.
[zb31 rg31]=wmdeepzb(mm,x31,y);
[zb32 rg32]=wmdeepzb(mm,x32,y);
[zb33 rg33]=wmdeepzb(mm,x33,y);
[zb34 rg34]=wmdeepzb(mm,x34,y);
[zb35 rg35]=wmdeepzb(mm,x35,y);
'Level 3 training done '

% x41,x42,x43 are inputs to the 3 fuzzy systems in Level 4.
x41(1:ntrain,1)=wmdeepyy(mm,zb31,rg31,x31);
x41(1:ntrain,2)=wmdeepyy(mm,zb32,rg32,x32);
x41(1:ntrain,3)=wmdeepyy(mm,zb33,rg33,x33);
x42(1:ntrain,1)=x41(:,2);
x42(1:ntrain,2)=x41(:,3);
x42(1:ntrain,3)=wmdeepyy(mm,zb34,rg34,x34);
x43(1:ntrain,1)=x41(:,3);
x43(1:ntrain,2)=x42(:,3);
x43(1:ntrain,3)=wmdeepyy(mm,zb35,rg35,x35);
% The following 3 lines train the 3 fuzzy systems in Level 4.
[zb41 rg41]=wmdeepzb(mm,x41,y);
[zb42 rg42]=wmdeepzb(mm,x42,y);
[zb43 rg43]=wmdeepzb(mm,x43,y);
'Level 4 training done '

% x51 is input to the fuzzy system in Level 5.
x51(1:ntrain,1)=wmdeepyy(mm,zb41,rg41,x41);
x51(1:ntrain,2)=wmdeepyy(mm,zb42,rg42,x42);
x51(1:ntrain,3)=wmdeepyy(mm,zb43,rg43,x43);
```

```
% Train the fuzzy system in Level 5.
[zb51 rg51]=wmdeepzb(mm,x51,y);
'Level 5 training done '
'Training done'

'Computing the training and testing errors'
% The meanings of the variables are the same as in training.
for i=1:N
    for j=1:3
        x11(i,j)=xx(i,j);
        x12(i,j)=xx(i,j+1);
        x13(i,j)=xx(i,j+2);
        x14(i,j)=xx(i,j+3);
        x15(i,j)=xx(i,j+4);
        x16(i,j)=xx(i,j+5);
        x17(i,j)=xx(i,j+6);
        x18(i,j)=xx(i,j+7);
        x19(i,j)=xx(i,j+8);
    end;
    y(i,1)=xx(i,12);
end;
x21(1:N,1)=wmdeepyy(mm,zb11,rg11,x11);
x21(1:N,2)=wmdeepyy(mm,zb12,rg12,x12);
x21(1:N,3)=wmdeepyy(mm,zb13,rg13,x13);
x22(1:N,1)=x21(:,2);
x22(1:N,2)=x21(:,3);
x22(1:N,3)=wmdeepyy(mm,zb14,rg14,x14);
x23(1:N,1)=x21(:,3);
x23(1:N,2)=x22(:,3);
x23(1:N,3)=wmdeepyy(mm,zb15,rg15,x15);
x24(1:N,1)=x22(:,3);
x24(1:N,2)=x23(:,3);
x24(1:N,3)=wmdeepyy(mm,zb16,rg16,x16);
x25(1:N,1)=x23(:,3);
x25(1:N,2)=x24(:,3);
x25(1:N,3)=wmdeepyy(mm,zb17,rg17,x17);
x26(1:N,1)=x24(:,3);
x26(1:N,2)=x25(:,3);
x26(1:N,3)=wmdeepyy(mm,zb18,rg18,x18);
x27(1:N,1)=x25(:,3);
x27(1:N,2)=x26(:,3);
x27(1:N,3)=wmdeepyy(mm,zb19,rg19,x19);
'Level 1 computing done'

x31(1:N,1)=wmdeepyy(mm,zb21,rg21,x21);
x31(1:N,2)=wmdeepyy(mm,zb22,rg22,x22);
x31(1:N,3)=wmdeepyy(mm,zb23,rg23,x23);
x32(1:N,1)=x31(:,2);
x32(1:N,2)=x31(:,3);
x32(1:N,3)=wmdeepyy(mm,zb24,rg24,x24);
x33(1:N,1)=x31(:,3);
x33(1:N,2)=x32(:,3);
x33(1:N,3)=wmdeepyy(mm,zb25,rg25,x25);
x34(1:N,1)=x32(:,3);
x34(1:N,2)=x33(:,3);
x34(1:N,3)=wmdeepyy(mm,zb26,rg26,x26);
x35(1:N,1)=x33(:,3);
x35(1:N,2)=x34(:,3);
x35(1:N,3)=wmdeepyy(mm,zb27,rg27,x27);
'Level 2 computing done'

x41(1:N,1)=wmdeepyy(mm,zb31,rg31,x31);
x41(1:N,2)=wmdeepyy(mm,zb32,rg32,x32);
x41(1:N,3)=wmdeepyy(mm,zb33,rg33,x33);
x42(1:N,1)=x41(:,2);
x42(1:N,2)=x41(:,3);
x42(1:N,3)=wmdeepyy(mm,zb34,rg34,x34);
x43(1:N,1)=x41(:,3);
x43(1:N,2)=x42(:,3);
x43(1:N,3)=wmdeepyy(mm,zb35,rg35,x35);
'Level 3 computing done'

x51(1:N,1)=wmdeepyy(mm,zb41,rg41,x41);
x51(1:N,2)=wmdeepyy(mm,zb42,rg42,x42);
x51(1:N,3)=wmdeepyy(mm,zb43,rg43,x43);
'Level 4 computing done'

yy=wmdeepyy(mm,zb51,rg51,x51); % yy is output of DCFS.
'Level 5 computing done'

% The following computes and plots the training and testing errors.
for i=1:N
    e(i)=y(i)-yy(i);
end;
err_train=0;
for i=1:ntrain
    err_train=err_train+e(i)^2;
end;
err_train=sqrt(err_train/ntrain);
err_test=0;
for i=ntrain+1:N
    err_test=err_test+e(i)^2;
end;
err_test=sqrt(err_test/(N-ntrain));
plot(e); % Plot the DCFS error curve in Fig. 12.
title(['DCFS: Training error = ' num2str(err_train) ', Testing error = '
num2str(err_test)]);
% End of main program.

function [zb ranges]=wmdeepzb(mm,xx,y)
% Train fuzzy system with input xx and output y to get zb and ranges,
% where zb is the c^{j1...jm} in FS (7) and ranges is the endpoints in (11).
extra=[xx,y];
[numSamples,m]=size(extra);
numInput=m-1;
for i=1:numInput
    fnCounts(i)=mm;
end;
ranges = zeros(numInput,2);
activFns = zeros(numInput,2);
activGrades = zeros(numInput,2);
searchPath = zeros(numInput,2);
numCells = 1; % number of regions (cells)
for i = 1:numInput
     ranges(i,1) = min(extra(:,i));
     ranges(i,2) = max(extra(:,i));
     numCells = numCells * fnCounts(i);
end;
baseCount(1)=1;
for i=2:numInput
    baseCount(i)=1;
    for j=2:i
        baseCount(i)=baseCount(i)*fnCounts(numInput-j+2);
    end;
end;
% Generate rules for cells covered by data
zb = zeros(1,numCells); % THEN part centers of generated rules
ym = zeros(1,numCells);
for k = 1:numSamples
    for i = 1:numInput
        numFns = fnCounts(i);
        nthActive = 1;
```

```
        for nthFn = 1:numFns
          grade = meb2(numFns,nthFn,extra(k,i),ranges(i,1),ranges(i,2));
          if grade > 0
              activFns(i,nthActive) = nthFn;
              activGrades(i,nthActive) = grade;
              nthActive = nthActive + 1;
          end;
        end;   % endfor nthFn
    end;  % endfor i
    for i=1:numInput
        if activGrades(i,1) >= activGrades(i,2)
            searchPath(i,1)=activFns(i,1);
            searchPath(i,2)=activGrades(i,1);
        else
            searchPath(i,1)=activFns(i,2);
            searchPath(i,2)=activGrades(i,2);
        end;
    end;
    indexcell=1;
    grade=1;
    for i=1:numInput
        grade=grade*searchPath(i,2);
indexcell=indexcell+(searchPath(numInput-i+1,1)-1)*baseCount(i);
    end;
    ym(indexcell)=ym(indexcell)+grade;
    zb(indexcell)=zb(indexcell)+extra(k,numInput+1)*grade;
end;  % endfor k
for j=1:numCells
    if ym(j) ~= 0
        zb(j)=zb(j)/ym(j);
    end;
end;
% Extrapolate the rules to all the cells
for i=1:numInput-1
    baseCount(i)=1;
    for j=1:i
        baseCount(i)=baseCount(i)*fnCounts(numInput-j+1);
    end;
end;
ct=1;
zbb = zeros(1,numCells);
ymm = zeros(1,numCells);
while ct > 0
    ct=0;
    for s=1:numCells
        if ym(s) == 0
            ct=ct+1;
        end;
    end;
    for s=1:numCells
        if ym(s) == 0
            s1=s;
            index = ones(1,numInput);
            for i=numInput-1:-1:1
                while s1 > baseCount(i)
                    s1=s1-baseCount(i);
                    index(numInput-i)=index(numInput-i)+1;
                end;
            end;
            index(numInput)=s1;
            zbnum=0;
            for i=1:numInput-1
                if index(i) > 1
                    zbb(s)=zbb(s)+zb(s-baseCount(numInput-i));
                    ymm(s)=ymm(s)+ym(s-baseCount(numInput-i));
                    zbnum=zbnum+sign(ym(s-baseCount(numInput-i)));
                end;
                if index(i) < fnCounts(i)
                    zbb(s)=zbb(s)+zb(s+baseCount(numInput-i));
                    ymm(s)=ymm(s)+ym(s+baseCount(numInput-i));
                    zbnum=zbnum+sign(ym(s+baseCount(numInput-i)));
                end;
            end;
            if index(numInput) > 1
                zbb(s)=zbb(s)+zb(s-1);
                ymm(s)=ymm(s)+ym(s-1);
                zbnum=zbnum+sign(ym(s-1));
            end;
            if index(numInput) < fnCounts(numInput)
                zbb(s)=zbb(s)+zb(s+1);
                ymm(s)=ymm(s)+ym(s+1);
                zbnum=zbnum+sign(ym(s+1));
            end;
            if zbnum >= 1
                zbb(s)=zbb(s)/zbnum;
                ymm(s)=ymm(s)/zbnum;
            end;
        end; % endif ym
    end; % endfor s
    for s=1:numCells
        if ym(s) == 0 & ymm(s) ~= 0
            zb(s)=zbb(s);
            ym(s)=ymm(s);
        end;
    end;
end; % endwhile ct

function yy=wmdeepyy(mm,zb,ranges,xx)
% Compute the output of fuzzy system with zb and ranges for input xx.
exapp=xx;
[numSamples,m]=size(exapp);
numInput=m;
for i=1:numInput
    fnCounts(i)=mm;
end;
activFns = zeros(numInput,2);
activGrades = zeros(numInput,2);
baseCount(1)=1;
for i=2:numInput
    baseCount(i)=1;
    for j=2:i
        baseCount(i)=baseCount(i)*fnCounts(numInput-j+2);
    end;
end;
for j=1:numInput
    for i1=1:2^j
        for i2=1:2^(numInput-j)
            ma(i2+(i1-1)*2^(numInput-j),j)=mod(i1-1,2)+1;
        end;
    end;
end;
e1sum=0;
for k = 1:numSamples
    for i = 1:numInput
        numFns = fnCounts(i);
        nthActive = 1;
        for nthFn = 1:numFns
grade = meb2(numFns,nthFn,exapp(k,i),ranges(i,1),ranges(i,2));
            if grade > 0
                activFns(i,nthActive) = nthFn;
                activGrades(i,nthActive) = grade;
                nthActive = nthActive + 1;
```

```
            end;
        end;
    end;
    for i=1:numInput
        nn(i,1)=activFns(i,1);
        nn(i,2)=nn(i,1)+1;
        if nn(i,1) == fnCounts(i)
            nn(i,2)=nn(i,1);
        end;
    end;
    a=0;
    b=0;
    for i=1:2^numInput
        indexcell=1;
        grade=1;
        for j=1:numInput
            grade=grade*activGrades(j,ma(i,j));
indexcell=indexcell+(nn(numInput-j+1,ma(i,numInput-j+1))-1)*base
Count(j);
        end;
        a=a+zb(indexcell)*grade;
        b=b+grade;
    end;
    yy(k)= a/b; % the fuzzy system output
end;  % endfor k

function y=meb2(n,i,x,xmin,xmax)
% Compute the value of the i'th membership function in Fig. 2 at x.
h=(xmax-xmin)/(n-1);
if i==1
    if x < xmin
        y=1;
    end;
    if x >= xmin & x < xmin+h
        y=(xmin-x+h)/h;
    end;
    if x >= xmin+h
        y=0;
    end;
end;
if i > 1 & i < n
    if x < xmin+(i-2)*h | x > xmin+i*h
        y=0;
    end;
    if x >= xmin+(i-2)*h & x < xmin+(i-1)*h
        y=(x-xmin-(i-2)*h)/h;
    end;
    if x >= xmin+(i-1)*h & x <= xmin+i*h
        y=(-x+xmin+i*h)/h;
    end;
end;
if i==n
    if x < xmax-h
        y=0;
    end;
    if x >= xmax-h & x < xmax
        y=(-xmax+x+h)/h;
    end;
    if x >= xmax
        y=1;
    end;
end;
```